\documentclass[usenatbib,usegraphicx]{mn2e}

\usepackage{longtable}
\usepackage{lscape}

\def\CIVdbl{{\rm C~}\kern 0.1em{\sc iv}~$\lambda\lambda 1548, 1550$}
\def\MgIIdbl{{\rm Mg~}\kern 0.1em{\sc ii}~$\lambda\lambda 2796, 2803$}
\def\NVdbl{{\rm N}\kern 0.1em{\sc v}~$\lambda\lambda 1238, 1242$}  
\def\OVIdbl{{\rm O}\kern 0.1em{\sc vi}~$\lambda\lambda 1031, 1037$}
\def\SiIVdbl{{\rm Si~}\kern 0.1em{\sc iv}~$\lambda\lambda1394, 1403$}
\def\AlIIIdbl{{\rm Al~}\kern 0.1em{\sc iii}~$\lambda\lambda1855,1863$}
\def\FeIIdbl{{\rm Fe~}\kern 0.1em{\sc ii}~$\lambda\lambda 2383, 2600$}

\def\AlII{\hbox{{\rm Al~}\kern 0.1em{\sc ii}}}
\def\AlIII{\hbox{{\rm Al~}\kern 0.1em{\sc iii}}}
\def\CaI{\hbox{{\rm Ca}\kern 0.1em{\sc i}}}
\def\CaII{\hbox{{\rm Ca}\kern 0.1em{\sc ii}}}
\def\CrII{\hbox{{\rm Cr}\kern 0.1em{\sc ii}}}
\def\C{\hbox{{\rm C~}}}
\def\CI{\hbox{{\rm C~}\kern 0.1em{\sc i}}}
\def\CII{\hbox{{\rm C~}\kern 0.1em{\sc ii}}}
\def\CIII{\hbox{{\rm C~}\kern 0.1em{\sc iii}}}
\def\CIV{\hbox{{\rm C~}\kern 0.1em{\sc iv}}}
\def\CV{\hbox{{\rm C}\kern 0.1em{\sc v}}}

\def\HI{\hbox{{\rm H~}\kern 0.1em{\sc i}}}
\def\HII{\hbox{{\rm H~}\kern 0.1em{\sc ii}}}
\def\Lya{\hbox{{\rm Ly}\kern 0.1em$\alpha$}}
\def\Lyb{\hbox{{\rm Ly}\kern 0.1em$\beta$}}
\def\Lyg{\hbox{{\rm Ly}\kern 0.1em$\gamma$}}
\def\Lyfive{\hbox{{\rm Ly}\kern 0.1em$5$}}
\def\Lysix{\hbox{{\rm Ly}\kern 0.1em$6$}}
\def\Lyseven{\hbox{{\rm Ly}\kern 0.1em$7$}}
\def\Lyeight{\hbox{{\rm Ly}\kern 0.1em$8$}}
\def\Lynine{\hbox{{\rm Ly}\kern 0.1em$9$}}
\def\Lyten{\hbox{{\rm Ly}\kern 0.1em$10$}}
\def\HeI{\hbox{{\rm He}\kern 0.1em{\sc i}}}
\def\HeII{\hbox{{\rm He}\kern 0.1em{\sc ii}}}
\def\FeI{\hbox{{\rm Fe~}\kern 0.1em{\sc i}}}
\def\FeII{\hbox{{\rm Fe~}\kern 0.1em{\sc ii}}}
\def\FeIII{\hbox{{\rm Fe~}\kern 0.1em{\sc iii}}}
\def\MnII{\hbox{{\rm Mn}\kern 0.1em{\sc ii}}}
\def\MgI{\hbox{{\rm Mg~}\kern 0.1em{\sc i}}}
\def\MgII{\hbox{{\rm Mg~}\kern 0.1em{\sc ii}}}
\def\MgIII{\hbox{{\rm Mg~}\kern 0.1em{\sc iii}}}
\def\MgIV{\hbox{{\rm Mg~}\kern 0.1em{\sc iv}}}
\def\NaI{\hbox{{\rm Na}\kern 0.1em{\sc i}}}
\def\NV{\hbox{{\rm N}\kern 0.1em{\sc v}}}
\def\NII{\hbox{{\rm N}\kern 0.1em{\sc ii}}}
\def\NIII{\hbox{{\rm N}\kern 0.1em{\sc iii}}}

\def\OVI{\hbox{{\rm O}\kern 0.1em{\sc vi}}}
\def\OIV{\hbox{{\rm O}\kern 0.1em{\sc iv}}}
\def\OI{\hbox{{\rm O}\kern 0.1em{\sc i}}}
\def\OII{\hbox{{\rm O}\kern 0.1em{\sc ii}}}
\def\OIII{\hbox{{\rm O}\kern 0.1em{\sc iii}}}
\def\PV{\hbox{{\rm P}\kern 0.1em{\sc v}}}
\def\SiII{\hbox{{\rm Si~}\kern 0.1em{\sc ii}}}
\def\SiIII{\hbox{{\rm Si~}\kern 0.1em{\sc iii}}}
\def\SiIV{\hbox{{\rm Si~}\kern 0.1em{\sc iv}}}
\def\SII{\hbox{{\rm S}\kern 0.1em{\sc ii}}}
\def\SIII{\hbox{{\rm S}\kern 0.1em{\sc iii}}}
\def\SIV{\hbox{{\rm S}\kern 0.1em{\sc iv}}}
\def\SVI{\hbox{{\rm S}\kern 0.1em{\sc vi}}}
\def\TiII{\hbox{{\rm Ti}\kern 0.1em{\sc ii}}}
\def\ZnII{\hbox{{\rm Zn}\kern 0.1em{\sc ii}}}
\def\kms{\hbox{km~s$^{-1}$}}

\def\a{$\alpha$ }

\def\lsim{\mathrel{\rlap{\lower4pt\hbox{\hskip1pt$\sim$}}
    \raise1pt\hbox{$<$}}}                
\def\gsim{\mathrel{\rlap{\lower4pt\hbox{\hskip1pt$\sim$}}
    \raise1pt\hbox{$>$}}}                

\title[]{The Extreme High-Velocity Outflow in Quasar PG0935+417}\label{sec:PG0935}
\author[]{Paola~Rodr\'iguez~Hidalgo$^{1}$, Fred Hamann$^{2}$, Patrick Hall$^{3}$ \\
$^{1}$Department of Astronomy and Astrophysics, Pennsylvania State University, University Park, PA 16802\\
$^{2}$Department of Astronomy, University of Florida, Gainesville, FL 32611\\
$^{3}$Department of Physics and Astronomy, York University, Toronto, ON, M3J 1P3, Canada }
\begin{document}

\pagerange{\pageref{firstpage}--\pageref{lastpage}} \pubyear{2002}

\maketitle

\label{firstpage}

\begin{abstract} 

We report the detection of \OVIdbl\ and \NVdbl\ absorption in a system of ``mini-broad" absorption lines (mini-BALs) previously reported to have variable \CIVdbl\ in the quasar PG0935+417. The formation of these lines in an extreme high-velocity quasar outflow (with $v \sim -50000$ $\kms$) is confirmed by the line variability, broad smooth absorption profiles, and partial covering of the background light source. \HI\ and lower ionization metals are not clearly present. The line profiles are complex and asymmetric, with Full Widths at Half Minimum (FWHMs) of different components in the range $\sim$660 to $\sim$2510 \kms. The resolved \OVI\, doublet indicates that these lines are moderately saturated, with the absorber covering $\sim$80\% of the quasar continuum source ($C_f \sim$0.8). We derive ionic column densities of order 10$^{15}$ cm$^{-2}$ in \CIV\ and several times larger in \OVI , indicating an ionization parameter of $\log U \gsim -0.5$. Assuming solar abundances, we estimate a total column density of $N_H \sim 5 \times 10^{19}$ cm$^{-2}$. 

Comparisons to data in the literature show that this outflow emerged sometime between 1982 when it was clearly not present \citep{Bechtold84} and 1993 when it was first detected \citep{Hamann97b}. Our examination of the \CIV\ data from 1993 to 2007 shows that there is variable complex absorption across a range of velocities from $-45000$ to $-54000$ \kms. There is no clear evidence for acceleration or deceleration of the outflow gas. The observed line variations are consistent with either changes in the ionization state of the gas or clouds crossing our lines of sight to the continuum source. If the former case, the recombination times constrain the location of outflow to be at a radial distance of $r \lsim$~1.2 kpc with density of $n_H \gsim 1.1 \times 10^{4}$ cm$^{-3}$. In the latter case, the nominal transit times of moving clouds indicate $r\lsim 0.9$ pc.

Outflows are common in Active Galactic Nuclei (AGN), but extreme speeds such as those reported here are extremely rare. It is not clear what properties of PG~0935+417 might produce this unusual outflow. The quasar is exceptionally luminous, with $L\sim6\times10^{47}$ ergs s$^{-1}$, but it has just a modest Eddington ratio, $L/L_{Edd}\sim 0.2$, and no apparent unusual properties compared to other quasars. In fact, PG~0935+417 has significantly less X-ray absorption than typical BAL quasars even though its outflow has a degree of ionization typical of BALs at speeds that are 2--3 times larger than most BALs. These results present a challenge to theoretical  models that invoke strong radiative shielding in the X-rays/far-UV to moderate the outflow ionization and thus enable its radiative acceleration to high speeds. 

\end{abstract}


\section{Introduction} 
\label{sec:1}

Outflows are fundamental constituents of Active Galactic Nuclei (AGN). They are commonly detected (e.g. \citealt{Crenshaw99}; \citealt{Reichard03}; \citealt{Hamann04}; \citealt{Trump06}; \citealt{Nestor08}; \citealt{Dunn08}; \citealt{Ganguly08} and references therein) and might be ubiquitous if the absorbing gas subtends only part of the sky as seen from the central continuum source. Indeed, the correlation between the black hole masses ($M_{BH}$) and the masses of the host galaxies ($M_{bulge}$ - \citealt{Gebhardt00}; \citealt{Merritt01}) implies a connection between the inner AGN and its surrounding host galaxy which could be partly explained by the ``feedback'' from AGN outflows (\citealt{Silk98}; \citealt{DiMatteo05}). AGN feedback might also play a role in blowing out the gas and dust from young galaxies and distributing metal-rich gas to the intergalactic medium.

\par One of the difficulties in understanding the outflows revealed by absorption features in AGN spectra is their tremendous diversity. Outflows observed via resonant line absorption of ions such as \CIVdbl\ have been previously classified based on their widths. Broad Absorption Lines (BALs), which show typical widths of several thousands of km s$^{-1}$, and outflowing Narrow Absorption Lines (NALs), with widths less than a few hundred km s$^{-1}$, are the most commonly studied classes of quasar outflow lines (\citealt{Weymann81}; \citealt{Turnshek84};  \citealt{Foltz86}; \citealt{Weymann91};  \citealt{Aldcroft94}; \citealt{Reichard03}; \citealt{Vestergaard03}; \citealt{Trump06}).
Absorption lines with intermediate widths, called ``mini-BALs", are just as common as BALs (Rodriguez Hidalgo et al., in prep.) and appear at a wide range of velocities, but only a handful of cases have been studied in detail (i.e., \citealt{Turnshek88}; \citealt{Januzzi96}; \citealt{Hamann97a}; \citealt{Churchill99}; \citealt{Yuan02}; \citealt{Narayanan04};  \citealt{Misawa07b}). Mini-BALs could represent a completely different type of absorber, or a particular line of sight through the same absorbers that we observe in other cases as BALs. A better characterization of the physical properties of these mini-BALs is therefore necessary to accomplish a more complete understanding of quasar outflows.

Of special interest are the cases with extremely large speeds because they present the biggest challenge to theoretical models of the acceleration (\citealt{Hamann02}; \citealt{Sabra03}). The few know cases of mini-BALs measured in the rest-frame UV at speeds approaching 0.2$c$ (see below) might be similar to the extreme high-velocity features reported in X-ray spectra of other quasars (\citealt{Chartas02}; \citealt{Pounds03}; \citealt{Reeves09}). 

Many current models of AGN outflows present the gas arising from the accretion disk. It remains unsettled, though, how these outflows are launched and driven. Several mechanisms and dynamical models have been proposed to explain their origin and how they are ejected from the inner quasar region: line radiation pressure alone (\citealt{Arav94}; \citealt{Murray95}; \citealt{Proga00}) or combined with magnetical forces (\citealt{deKool95}; \citealt{Everett05}; \citealt{Proga04}). Radiation pressure seems to play a dominant role as it explains the relation between the AGN luminosity and the terminal velocity of the outflow (\citealt{Laor02}). It is also supported by the evidence for ``line locking" in a few cases (\citealt{Turnshek88}; \citealt{Srianand02}; \citealt{Hamann10}). 

\par \citet{Hamann97b} reported the presence of a \CIVdbl\ mini-BAL outflow system (FWHM of the whole profile $\sim$ 1500 km s$^{-1}$) at a velocity shift of 
$v\sim -50000$ \kms\ in the $z_{em}\sim 1.96$ quasar PG~0935+417. This bright quasar ($V$= 16.2) is radio-quiet and has an emission-line redshift of z$_{em}$=1.966 (\citealt{Tytler92}; \citealt{Hewitt93}). Its spectrum characterizes it as a non-BALQSO (zero Balnicity Index, as defined in \citealt{Weymann91}) because the outflow velocity is large and the velocity width is small. Several methods have been used to confirm the outflow nature of the \CIV\, absorber. In particular, Keck observations at resolution $\sim$7 km s$^{-1}$ confirmed that the mini-BAL profile is ``smooth'' (i.e., it is not composed of many narrow lines) and therefore the large FWHM is likely to have an outflow origin (\citealt{Hamann97b}). Also, \citet{Narayanan04} reported variability in the \CIV\ mini-BAL on time scales as short as $\sim$1 year in the quasar rest-frame. We report for the first time here that a spectrum of PG~0935+417 obtained in 1982 (\citealt{Bechtold84}) reveals the absence of the \CIV\ mini-BAL at a high significance. Therefore, this outflow system emerged sometime between 1982 and its first reported measurement in 1993 (see \S\ref{obs}). This is now just the fourth example of observations catching an emerging AGN outflow (see also \citealt{Ma02}; \citealt{Hamann08}; \citealt{Leighly09}). 

The velocity of the PG0935+417 mini-BAL is the second highest flow speed found in the UV spectrum of a non-BALQSO. The discovery by \citet{Januzzi96} of \CIV, \NV\, and \OVI\, outflowing at $-56000$ km s$^{-1}$ in another luminous quasar, PG2302+029, is the highest. Although \citet{Januzzi96} speculated that the broad-ish absorption lines in PG2302+029, with full widths at half minimum (FWHMs) between 3000 and 5000 km s$^{-1}$, might be cosmologically intervening instead of intrinsic to the quasar (for example, due to warm intra-cluster gas, see \citealt{Januzzi96}), more recent observations have shown that these features have variable strengths (over a couple of years time scale, see \citealt{Januzzi10}), implying that they form in a dynamic quasar outflow. 

\par In this paper we present new measurements of \OVIdbl\ and \NVdbl\ in the PG~0935+417 outflow, measured first in \CIV\ only by \citet{Hamann97b}, using archival Hubble Space Telescope (HST) spectra. We include a search of other ions in the same outflow to place limits on the degree of ionization and total column density in this outflow. We also expand the variability study of the \CIV\ mini-BAL by using more recent spectra from the Sloan Digital Sky Survey (SDSS - 2003) and spectra we obtained at the Kitt Peak National Observatory (KPNO - 2007), which enlarges the original rest-frame measurement interval from 2 to almost 5 years. We defer discussion of other narrower \CIV\ absorption systems in the PG~0935+417 spectrum to a future paper (\citealt{Hamann10c}) that includes higher resolution spectroscopy.

\section{Data}
\label{obs}

\begin{figure*}
 \begin{minipage}{180mm}
 \begin{center}
\includegraphics[width=19cm]{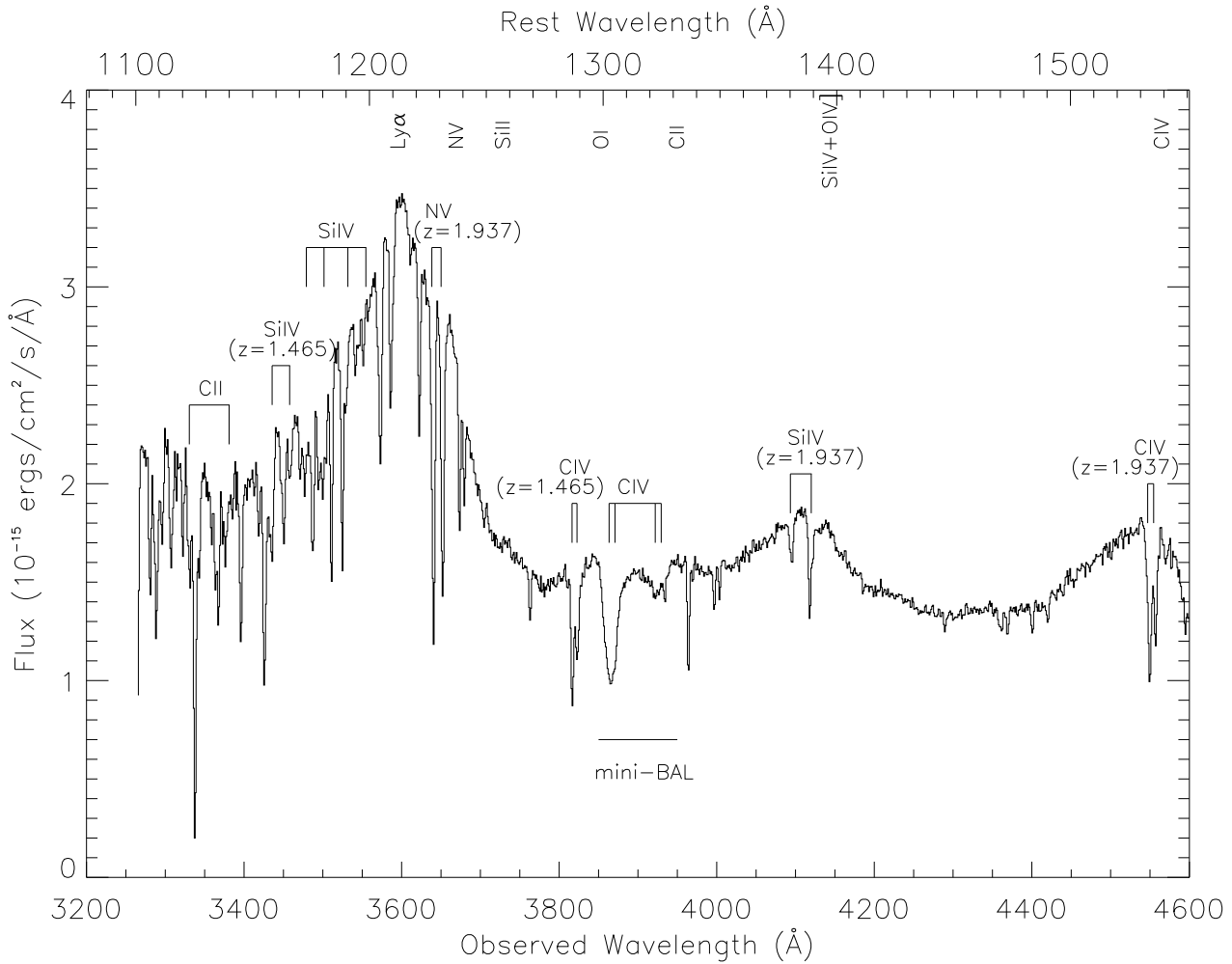}
\caption[Optical spectrum obtained at Lick observatory in 1996]{Optical spectrum obtained at Lick observatory in 1996. Prominent broad emission lines are labeled across the top. Tick marks indicate the location of absorption lines. The \CIV\, mini-BALs lie at an observed wavelength of $\sim$ 3865 \AA\, ($z_{abs} \approx$ 1.496) and $\sim$ 3925 \AA\, ($z_{abs} \approx$ 1.534). We have marked the location of \SiIV\, and \CII, in the same outflow, although we do not confirm their detection. Besides that system, we have labeled associated systems at $z_{abs} =$ 1.937 and at $z_{abs} =$ 1.465, which are used to reject detections of other ions in the absorption system of interest that could be confused with ions in these systems. Rest wavelength is defined based on the value of $z_{em}\approx$~1.966 given in \citet{Tytler92}.}
\label{LICKspec}
 \end{center}
 \end{minipage}
\end{figure*}

\par Table \ref{obslog} summarizes the PG0935+417 data used in this study. We analyzed spectra previously obtained during four observing runs (from 1993 to 1999) using the KAST spectrograph at the 3.0 m telescope at the University of California Observatories (UCO) Lick Observatory, with the wavelength coverage and resolutions (near those wavelengths of interest) shown in Table \ref{obslog}. See Narayanan et al. (2004) for more information on these observations and data reductions. We verified the wavelength calibrations of these data by using spectra with resolution R $= \lambda / \Delta\lambda \approx$ 34000 ($\sim$ 0.13 \AA\, or $\sim$ 9 km s$^{-1}$) obtained with HIRES at the 10.0 m telescope of the W. H. Keck observatory on January 1998 (\citealt{Hamann10c}), already reduced and shifted to vacuum in the heliocentric frame. Comparison to intervening absorption lines (\FeII\, $\lambda$2383 and \MgII\, $\lambda$2796, $\lambda$2804) in the Keck spectra suggested a slight shift of $\sim$~1 \AA\, of the Lick spectra to match these narrow absorption lines. The Keck spectra are not used further in this work due to the added uncertainty in the flux calibrations across the absorption feature.

\begin{figure*}
\begin{minipage}{180mm}
\begin{center}
\includegraphics[width=11cm,angle=90]{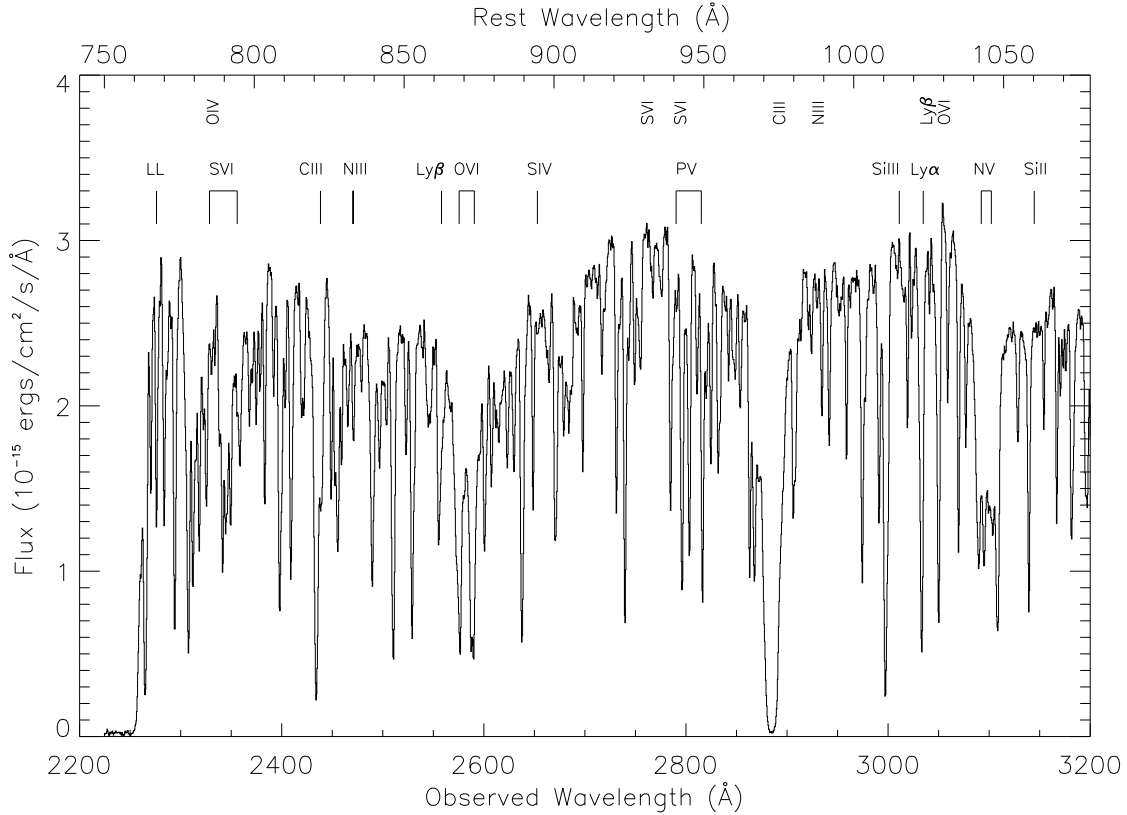}
\caption[Ultraviolet spectrum from the HST archive]{Ultraviolet spectrum from the HST archive (1994 and 1995 spectra combined). Possible broad emission lines are labeled across the top. Tick marks indicate the location of lines that might be present at the $z_{abs} \approx$ 1.496 location of interest. The strong absorption feature at observed wavelength $\sim$ 2870 \AA\, is an unrelated Damped Ly$\alpha$ at $z_{abs} \sim$1.374. The strong Lyman Limit (LL) at an observed wavelength $\sim$ 2250 \AA\, corresponds to the also unrelated absorption system at $z_{abs} \sim$ 1.465.}
\label{HSTspec}
\end{center}
\end{minipage}
\end{figure*}

\begin{table}
\caption{Observation logs \label{obslog}}
\begin{tabular}{crrrr}
\hline
\noalign{\vskip 4pt}
Observatory & Date & $\lambda$ Range & Exp. time & $\Delta_v$  \\
 & &  (\AA) & (s) & (km/s)  \\
\hline
\noalign{\vskip 4pt}
Lick & 1993 Mar$^1$ & 3250-5350 & 1200 & 375$^1$ \\ 
\noalign{\vskip 4pt}
HST & 1994 Oct & 2270-3270 & 2046 & 230 \\ 
\noalign{\vskip 4pt}
HST & 1995 Nov & 2270-3270 & 8910 & 230\\
\noalign{\vskip 4pt}
Lick & 1996 Mar & 3250-4600 & 2700 & 230\\ 
\noalign{\vskip 4pt}
Lick & 1997 Feb & 3250-6000 & 2700 & 230\\ 
\noalign{\vskip 4pt}
Lick & 1999 Jan & 3250-6000 & 3000 & 230\\ 
\noalign{\vskip 4pt}
SDSS & 2003 Jan & 3800-9200 & 2250 & 150 \\ 
\noalign{\vskip 4pt}
KPNO & 2007 Jan & 3600-6200 & 9000 & 200\\
\hline
\end{tabular}
\hspace{1.5cm} $^1$ Date and resolution were incorrect in Narayanan et al. (2004).
\end{table}

\par To look for the presence of other ions at the same redshift as the \CIV\, mini-BAL, we examined archival Hubble Space Telescope spectra obtained with the Faint Object Spectrograph (FOS) using the G270H grating and a 0.3$^{\prime\prime}$ aperture on 1994 October 7 and 1995 November 13. Four exposures in 1995 and one in 1994 provided total exposure times of 8910 and 2046 s, respectively. These spectra have a resolution R = 1300 (230 km s$^{-1}$) and were obtained from the Space Telescope Science Institute archives already reduced and calibrated. We also verified the wavelength calibration using Galactic absorption lines (\FeII\, $\lambda$2383, $\lambda$2600, and \MgII\, $\lambda$2796, $\lambda$2804), which we assumed to be at their laboratory wavelengths. The absorption features measured with HST did not vary significantly (beyond the noise) between the 1994 and 1995 observations. We therefore combine all of these spectra to improve the signal-to-noise ratio and, hereafter, discuss only the combined HST spectrum.

The HST data were not taken simultaneously with the Lick data (see Table~\ref{obslog}). The two Lick spectra obtained closest in time to the HST observations are from 1993 and 1996 (Table~\ref{obslog}). In our analysis that combines the HST and Lick measurements, we consider primarily the Lick 1996 spectrum because i) it is temporally closest to the HST data, and ii) the \CIV\ mini-BAL profile measured in 1996 is roughly similar to the \OVI\ mini-BAL profile measured with HST. We note, however, that the \CIV\ mini-BAL did vary signficantly between 1993 and 1996 (see \S\ref{var_ana}) and therefore some unknown amount of variation might have occurred between the HST observations in 1994/95 and the Lick measurement in 1996.

\par Finally, we examined more recent spectra to extend the time baseline and monitor the variability of the \CIV\, mini-BAL absorption line (see Table \ref{obslog}). Archival spectra were obtained from the Sloan Digital Sky Survey (SDSS) with resolution R $\sim$ 2000 (150 km s$^{-1}$; \citealt{Adelman-McCarthy08}). We also obtained more recent spectra with the GoldCam spectrometer at the 2.1 m telescope at the Kitt Peak National Observatory (KPNO). We used the 26new grating to obtain a resolution $R \sim 1500$ (200 km s$^{-1}$), similar to the other resolutions in this study, and sufficient to resolve the \CIV\, mini-BAL system.

The KPNO data were reduced and spectra were extracted using packages from the Image Reduction and Analysis Facility (IRAF) and our own software (coded in IDL). HeNeAr lamps were used for the wavelength calibrations and quartz lamps were used for the flat-fields, both located inside the spectrograph. We used the overscan region on the CCD to subtract the bias. A one dimensional response function was sufficient to create an appropriate flat field. Relative flux calibrations were performed using KPNO flux standards observed on the same night.

\section{Analysis}
\label{ana}

\subsection{Line Identification}
\label{identf}

Figure \ref{LICKspec} shows the 1996 optical spectrum of PG0935+417 with several absorption line  systems labeled. The rest-frame wavelengths here and elsewhere in this paper are defined relative to the redshift $z_{em} =$~1.966 from \citet{Tytler92}, which is based on low-ionization broad emission lines. Thus, some amount of blueshift in the higher ionization emission lines, such as \CIV\ in Figure 1, can be expected relative to this value of $z_{em}$ (\citealt{Shen07}). 

The \CIV\, mini-BAL has a complex, time-variable absorption profile (\citealt{Hamann97b}, \citealt{Narayanan04}, \S\ref{var_ana} below). In the 1996 spectrum shown in Figure \ref{LICKspec}, the strongest absorption component at $z_{abs}\approx$~1.496 ($v\approx$~51260 \kms) is accompanied by a distinct  weaker feature centered at roughly $z_{abs}\approx$~1.534 ($v\approx$~46820 \kms). There are also two much narrower \CIV\, absorption systems ($z_{abs} \approx$ 1.465, $z_{abs} \approx$ 1.937). The latter is an ``associated'' system already studied in \citet{Hamann97b} and discussed further by \citet{Hamann10c}.  The former is likely to be an unrelated intervening system because it shows a lack of variability (see \S\ref{var_ana}) and its profile at high resolution (in the Keck HIRES spectrum) shows a narrow multi-component profile.

We searched the HST spectrum for other lines at the redshift of the \CIV\, mini-BAL. Figure \ref{HSTspec} shows the combined HST spectrum with the locations of lines typically found in BAL systems marked at the mini-BAL redshift of $z_{abs}=$~1.496 (see e.g., BAL spectra in \citealt{Hamann98}; \citealt{Arav01}; \citealt{Leighly09}). We detect for the first time \OVIdbl\ and \NVdbl\ mini-BALs in the HST spectra at velocities very similar to the \CIV\, mini-BAL originally reported by \citet{Hamann97b}. We searched for other lines at this same redshift, including, Ly$\alpha$ $\lambda$1215, Ly$\beta$ $\lambda$1026, \OI\, $\lambda\lambda$989,1302, \CII\, $\lambda$1037, 1335, \CIII\, $\lambda$977, \NII\, $\lambda$1084, \NIII\, $\lambda$990, \SiII\, $\lambda\lambda\lambda$1190,1193,1260, \SiIII\, $\lambda$1207, \SiIV\, $\lambda\lambda$1394,1403,  \PV\, $\lambda\lambda$1118,1128, \SIV\, $\lambda$1063, and \SVI\, $\lambda\lambda$ 933,946. No other detections are surely confirmed, partly due to the difficulty that the Ly$\alpha$ forest imposes, and partly due to the poor coincidence in wavelength with other ions of the other systems mentioned above. In several cases useful for our analysis below (\CIII, \NIII, \PV, \SiIV, Ly$\alpha$, and Ly$\beta$) we place upper limits on the absorption line strengths (see \S\ref{fits}). The strong absorption feature at an observed wavelength of $\sim$2885 \AA\, is an unrelated Damped Ly$\alpha$ line at $z_{abs} =$ 1.37 (\citealt{Lanzetta95}; \citealt{Turnshek02}). The Lyman limit at $\sim$ 2240 \AA\, (Figure \ref{HSTspec}) corresponds to the narrow absorption system at $z_{abs} \approx$ 1.465. There is no significant Lyman limit absorption related to the mini-BAL outflow system at $z_{abs} \approx$ 1.496.

\begin{figure}
\begin{center}
\includegraphics[width=9cm]{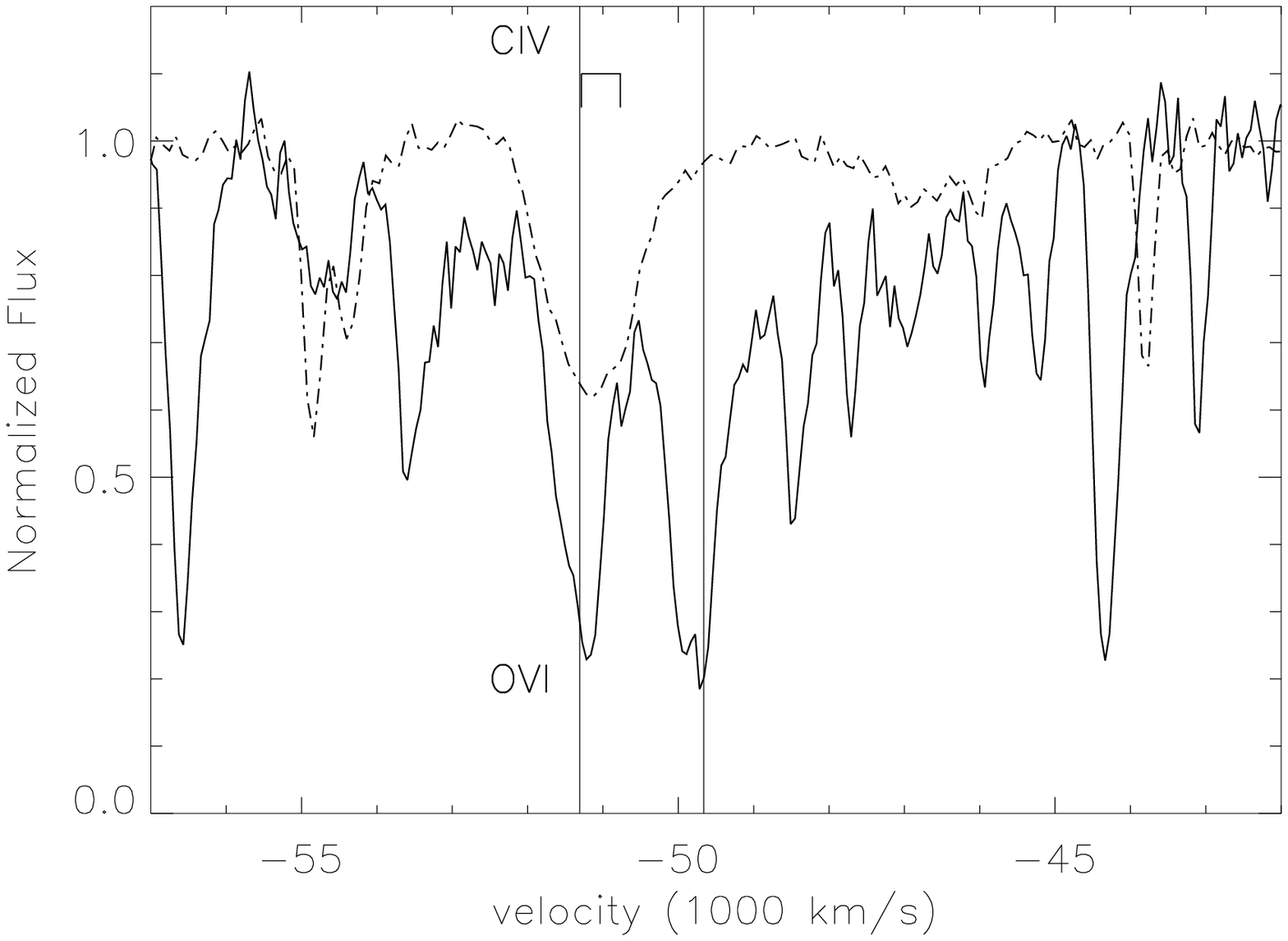}
\includegraphics[width=9cm]{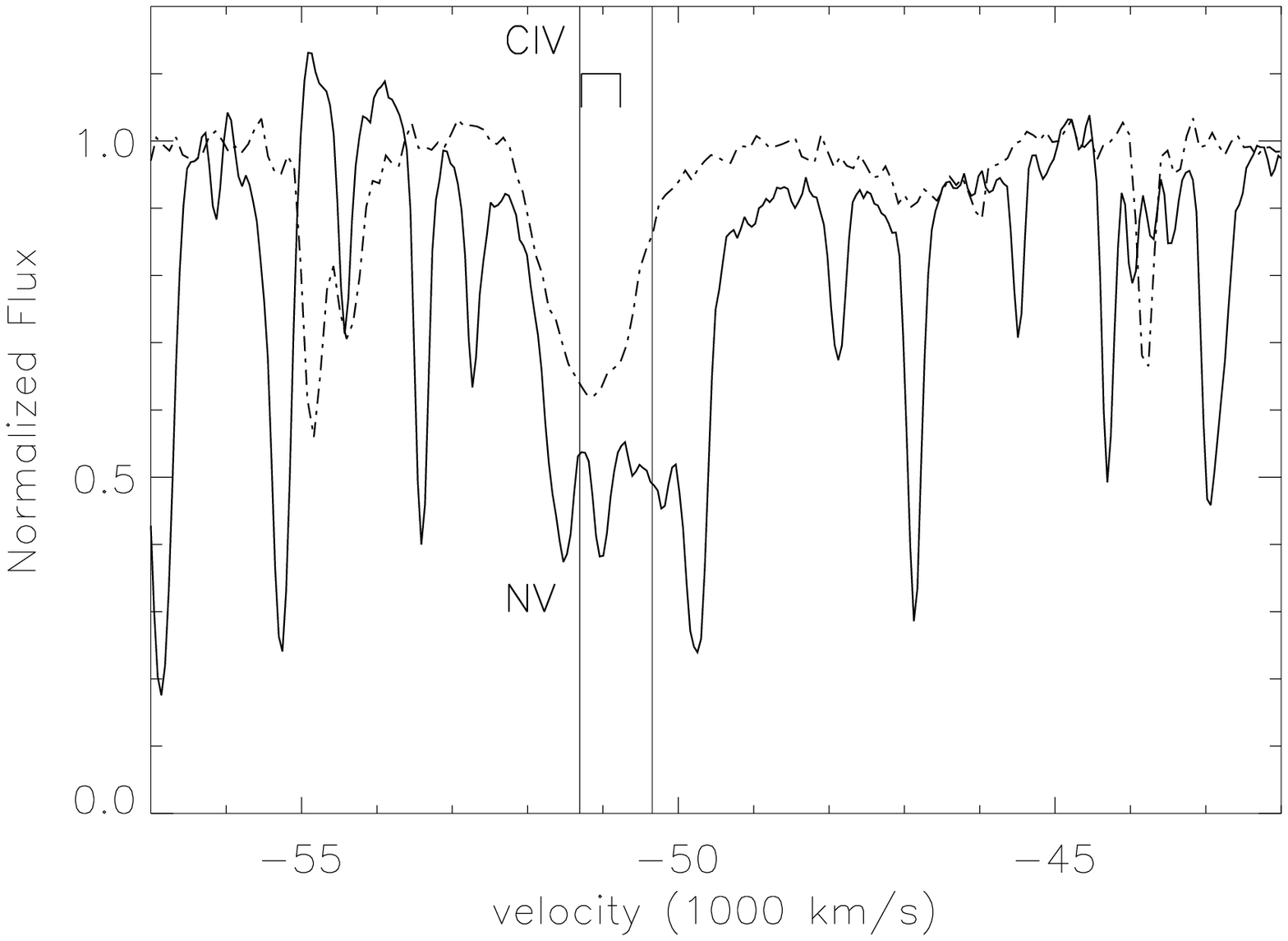}
\caption[HST normalized spectra]{Normalized HST spectra showing the high-velocity mini-BALs for \OVI\, and \NV\, (solid lines), with over-plotted \CIV\, from the normalized Lick spectrum (dotted-dashed lines). The brackets above the spectra indicate the \CIV\, doublet locations, while the vertical lines show the expected positions of the \OVI\, and \NV\, doublets at the same velocity as \CIV. The velocity scale corresponds to the short wavelength components of the doublets relative to the emission redshift. The HST spectrum includes many lines unrelated to the mini-BALs (Ly$\alpha$ forest). In the case of \NV, although strong absorption is present, the doublet appears to be blended with Ly$\alpha$ forest lines.}
\label{2panel}
\end{center}
\end{figure}

Figure \ref{2panel} shows the \CIV\, mini-BAL in Lick 1996 spectrum (dot-dashed curve) plotted over the \NV\, and \OVI\, absorption measured with HST (solid curve) on a velocity scale relative to the quasar emission redshift. The two vertical lines in each panel represent the \OVI\, and \NV\, doublet positions at the redshift $z_{abs} \approx$~1.496 of the deepest part of \CIV\, mini-BAL. Although the Ly$\alpha$ forest contaminates the \OVI\, and \NV\ measurements, the correct doublet separation in \OVI\ and the strong and broad absorption evident in both lines at the same velocity as the \CIV\, mini-BAL clearly indicates \OVI\ and \NV\ mini-BALs are both present (see also Figure \ref{fit1} and our line profile fits in \S\ref{fits} below). There is also evidence, e.g., in \OVI , for absorption related to the weaker and lower velocity component of the \CIV\ mini-BAL at $z_{abs}\approx$~1.534 and $v \sim -47000$ \kms.

\subsection{Continuum Normalization}
\label{cont_norm}

\par
\begin{figure}
\begin{center}
\includegraphics[width=9cm]{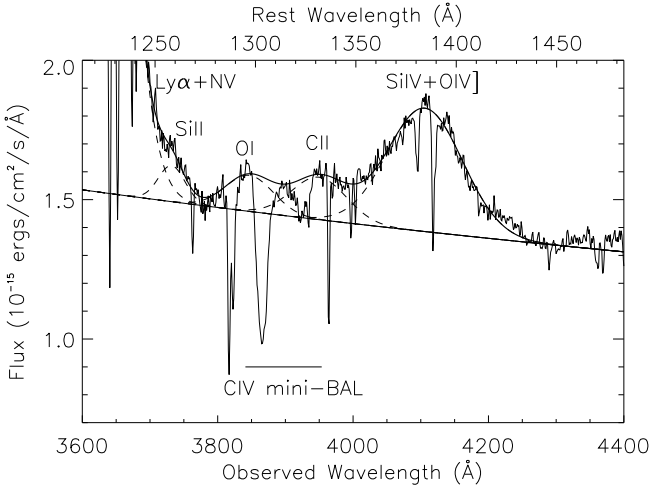}
\caption[Normalization of the Lick 1996 spectrum]{Normalization of the Lick 1996 spectrum. The solid straight line under the spectrum represents a fit to the continuum with a power law. Dashed lines represent our fit to Gaussians for the following emission lines: Ly$\alpha$+\NV\, (right slope), \SiII, \OI, \CII, and \SiIV+\OIV] around the \CIV\, mini-BAL, marked with an horizontal line under the spectrum. Because the absorption feature at the observed $\lambda \sim$ 3925 \AA\, appears to be real absorption it was excluded from the emission fitting.}
\label{96norm}
\end{center}
\end{figure}

 To facilitate measurements of the absorption lines, we normalized the Lick, HST, KPNO and SDSS spectra by fitting pseudo-continua across the tops of the absorption features. In Figure \ref{96norm} we show the normalization of the Lick 1996 spectrum. Our goal of measuring the absorption lines requires us to extrapolate a pseudo-continuum across the top of the \CIV\ mini-BAL. This goal is complicated by the fact that the absortion is fairly broad and there is weak underlying emission from \OI\, $\lambda$1305\AA\, and \CII\, $\lambda$1335\AA. Thus the normalization was achieved in two steps. First, we fit a power law to the continuum (resulting in $f_{\lambda} \propto \lambda^{-0.8}$), constrained in three narrow wavelength bands free of emission or absorption lines available in this spectrum (1272-1276 \AA, 1434-1443 \AA, and 1449-1455 \AA, in the rest frame of the quasar). Second, we fit single Gaussians to all the emission lines present around the \CIV\, mini-BAL (e.g., \SiII\, $\lambda$1263\AA, \OI\, $\lambda$1305\AA, \CII\, $\lambda$1335\AA, and \SiIV+\OIV] $\lambda$1400\AA). The location of this pseudo-continuum is particularly uncertain in the rest-frame region around $\sim$1280-1350 \AA\, due to the several parameters that take part in the fitting of the weaker emission lines (\SiII, \OI\, and \CII). These emission lines were fitted through an interactive process: we started the fit allowing their Full Width at Half Maxima (FWHM) and centroid $z_{em}$ to be free parameters, and corrected and fixed these parameters as we inspected the results of the fits, although the final parameters did not deviate largely from the original inputs. We masked the absorption features and constrained our fits to these emission lines to be redshifted with respect to the redshift of the stronger \SiIV+\OIV] feature, for which the best single-Gaussian fit yields $z_{em} =$ 1.939. Our final fits suggested a shift by $\sim$~300 km s$^{-1}$; this agrees well with the value $z_{em}$ = 1.966 reported by \citet{Tytler92}, which is also based on low-ionization lines (\MgII\, and Ly$\alpha$), and with recent estimates of the shifts between the \SIV\, and \MgII\, emission lines (\citealt{Shen07}). We constrained the FWHM of \OI\, and \CII\, to be narrower than the \SiIV+\OIV] FWHM, obtaining a FWHM of roughly 60\% of the \SiIV+\OIV] FWHM, which agrees with the results of extensive fitting of quasar composite spectra (Craig Warner, private communication). Finally, a weak additional Gaussian was used to fit the red wing of the Ly$\alpha$+\NV$\lambda$1240 \AA\, blend. We would like to note that the expected redshifts and FWHMs of the emission lines are used only as a guide to obtain a good fit to the unabsorbed spectrum.

\begin{figure}
\begin{center}
\includegraphics[height=6.5cm]{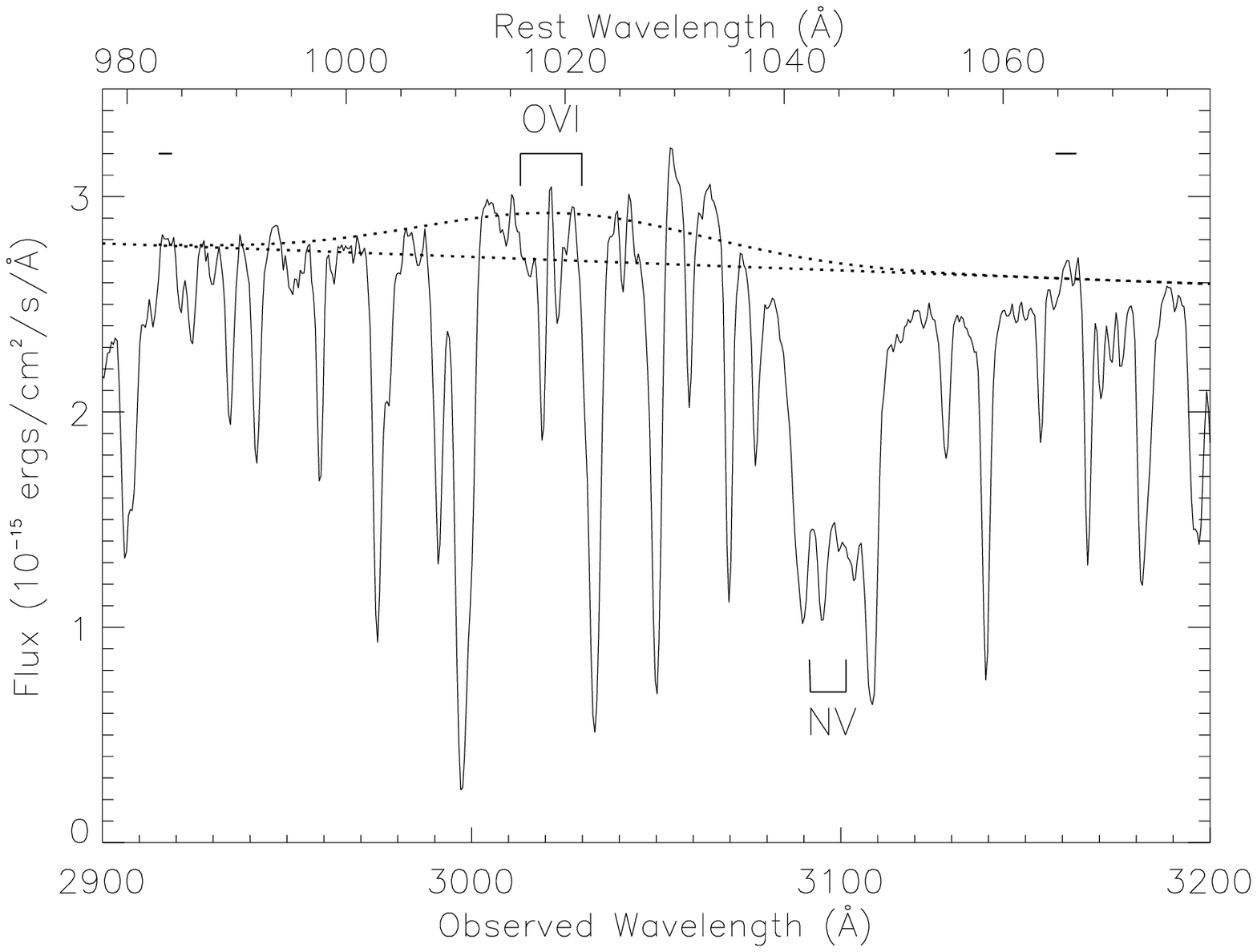}
\includegraphics[height=6.5cm]{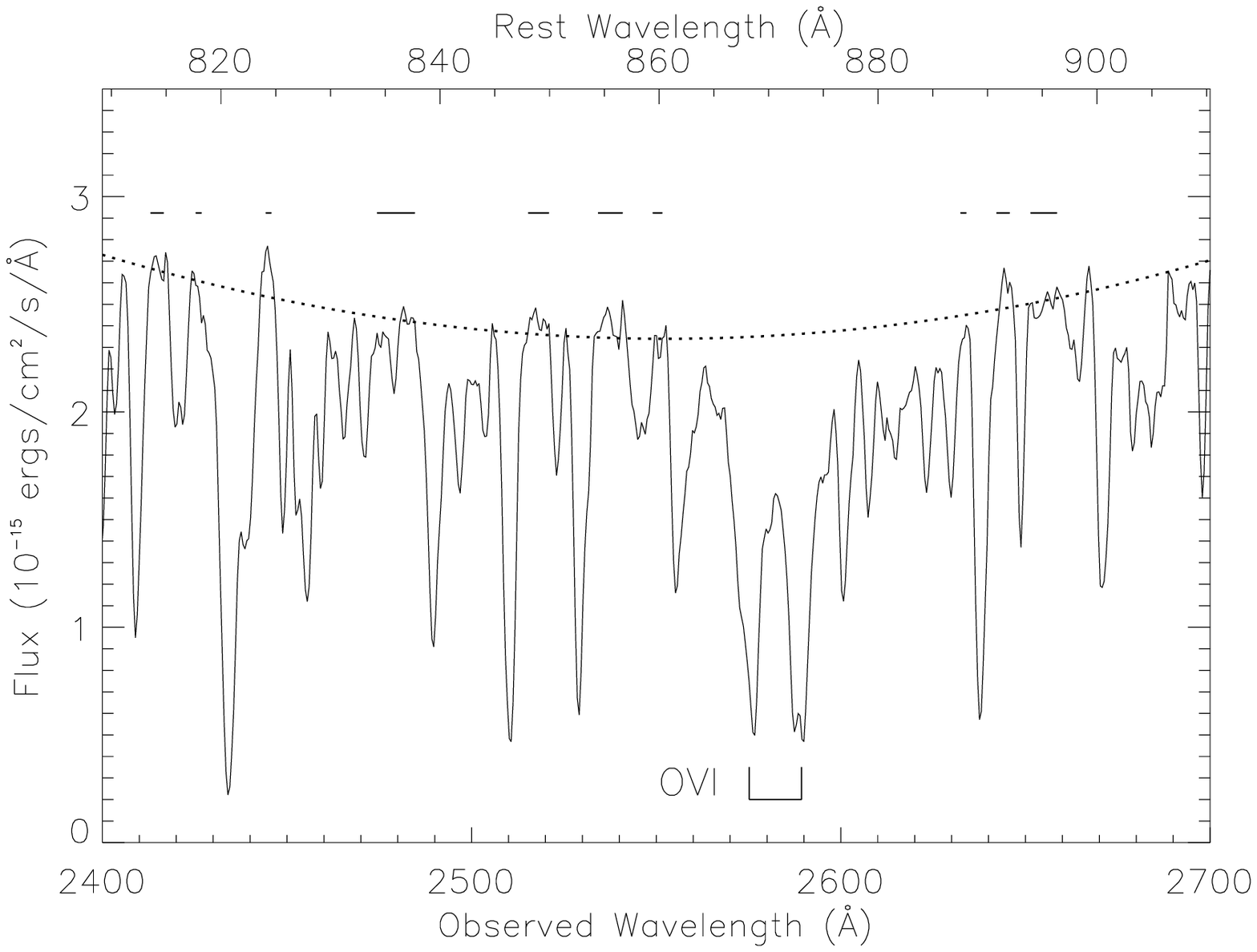}
\caption[Normalization of the HST spectrum]{Normalization of the HST spectrum (two epochs combined) around the \NV\, $\lambda\lambda$1239,1243 absorption line
($\sim$1035-1065 \AA, in the quasar rest frame) -- top -- and around the \OVI\, $\lambda\lambda$1032,1038 absorption line
($\sim$860-890 \AA) -- bottom. Horizontal solid lines above each spectrum represent the ranges chosen for the polynomial fit,
which is represented by dotted lines. The continuum at the left of \NV\, (top) also shows a \OVI\, broad emission line, represented as a single Gaussian and extracted from this spectrum as well.}
\label{HSTnorm}
\end{center}
\end{figure}

Figure \ref{HSTnorm} shows our fits to the continuum and emission lines in the HST spectra around the \NV\, (top) and \OVI\, (bottom) absorbers. The Ly$\alpha$ forest makes it difficult to define a continuum in these regions. In order to exclude the numerous unrelated absorption lines in the intervening Ly$\alpha$ forest, we fit the continua locally around the absorption of interest and constrained the fits (dotted lines) using only very narrow ranges in wavelength (indicated by the solid horizontal lines above the spectra). Our approach is to i) use the lowest order polynomial possible to avoid over-fitting undulations in the spectrum that might not be real, and ii) adopt a conservatively low continuum height to avoid overestimating the mini-BAL absorption strength, especially if there is absorption in very broad wings. In the case of the \OVI\, absorption, we fit the local continuum with a second order polynomial. For the \NV\, absorption (top panel of Figure \ref{HSTnorm}), we fit a straight line to the continuum and add a single Gaussian to represent the \OVI\, broad emission line. The location and shape of this Gaussian is again constrained by narrow wavelength ranges not absorbed by the Ly$\alpha$ forest. We centered the \OVI\, emission Gaussian at a redshift z$_{em} \approx$ 1.920, which is consistent with the position of the \CIV\ emission line shown in Figure 1 and with the adopted redshift of $z_{em}\approx 1.966$ if we consider the usual trend for decreasing redshifts in higher ionization emission lines (see above). The FWHM of this Gaussian \OVI\, emission line is  $\sim$~9000 km s$^{-1}$, consistent with our measurement of the \CIV\, emission line (FWHM(\CIV) = 9930 km s$^{-1}$). Although, as Figure \ref{HSTnorm} shows, our final fit to the \OVI\, emission line could be stronger and wider in the red wing, we prefer to again adopt a conservative approach using a weak and slightly narrower \OVI\, emission line to avoid overestimating the strength of the \NV\, mini-BAL.

Similar local continua were drawn and used to normalize the HST spectrum around the wavelengths corresponding to \SiIV, and \PV\, in the mini-BAL system, where low order polynomial functions were sufficient. \CIII, \NIII, Ly$\alpha$, and Ly$\beta$ lie in regions already normalized for the \OVI\, and \NV\, absorption (see Figures \ref{HSTspec} and \ref{HSTnorm}).

\subsection{Line Measurements and Physical Quantities}
\label{fits}

\par The high-resolution spectrum across the \CIV\, mini-BAL (\citealt{Hamann97b}) indicates that these features are truly broad and smooth and fully resolved in the Lick and HST spectra discussed here. Thus, we assume the measured line intensities at each velocity, $v$, are given by

\begin{equation}
I_v = I_o (1-C_f)+C_f I_o e^{-\tau_v}
\label{Ieq}
\end{equation}

\noindent where $I_o$ is the intensity of the continuum, $C_f$ is the line-of-sight coverage fraction (0 $\leq C_f \leq 1$), and $\tau_v$ is the line optical depth (see \citealt{Hamann99} for more information on coverage fraction). For simplicity, and given the blending in the \CIV\, doublet and the confusion caused by blends with Ly$\alpha$ forest lines in the \OVI\, and \NV\, mini-BALs, we consider only single values of $C_f$ across the line profiles. We also adopt a fitting scheme based on gaussian optical depth components defined by

\begin{equation}
\tau_v = \tau_o\, e^{-\frac{\Delta v^2}{b^2}}
\label{taueq}
\end{equation}

\noindent where $\tau_o$ is the line center optical depth, $\Delta v$ is the velocity shift from line center and $b$ is the Doppler parameter. We note that the velocity field in the absorption line gas is more complex than one or several gaussian $\tau_v$ distributions. This is evident, for example, from the complex absorption profiles in Figure \ref{variab} and the discussion of the mini-BAL variability in \S\ref{var_ana} below. Nonetheless, fits with simple gaussians allow us to characterize the line strengths, widths and velocity shifts in a way that takes into account the blended doublets and enables comparisons between \CIV, \NV\, and \OVI\, with their different doublet separations. 

\begin{figure}
\begin{center}
   \includegraphics[width=9cm]{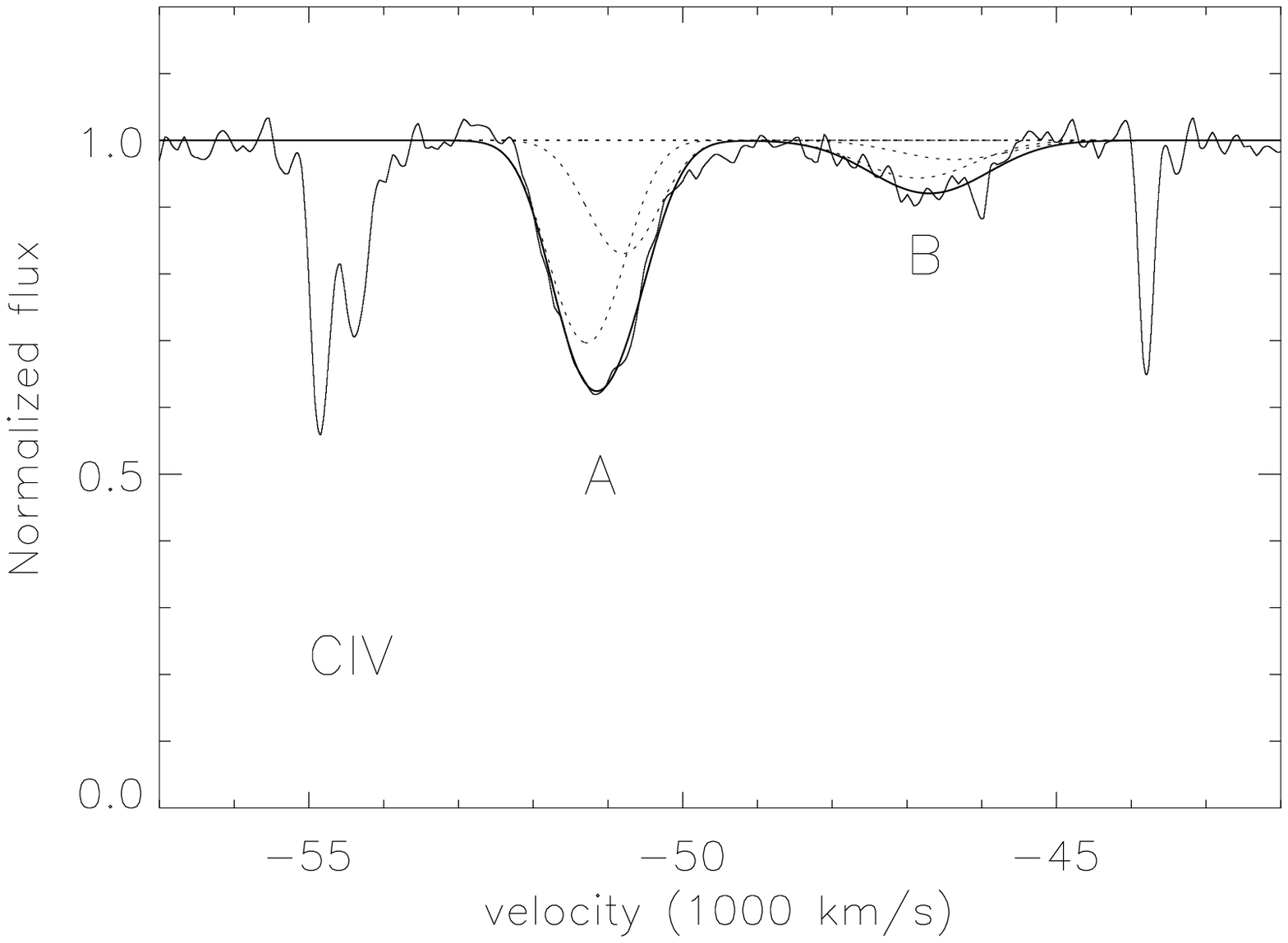}
   \includegraphics[width=9cm]{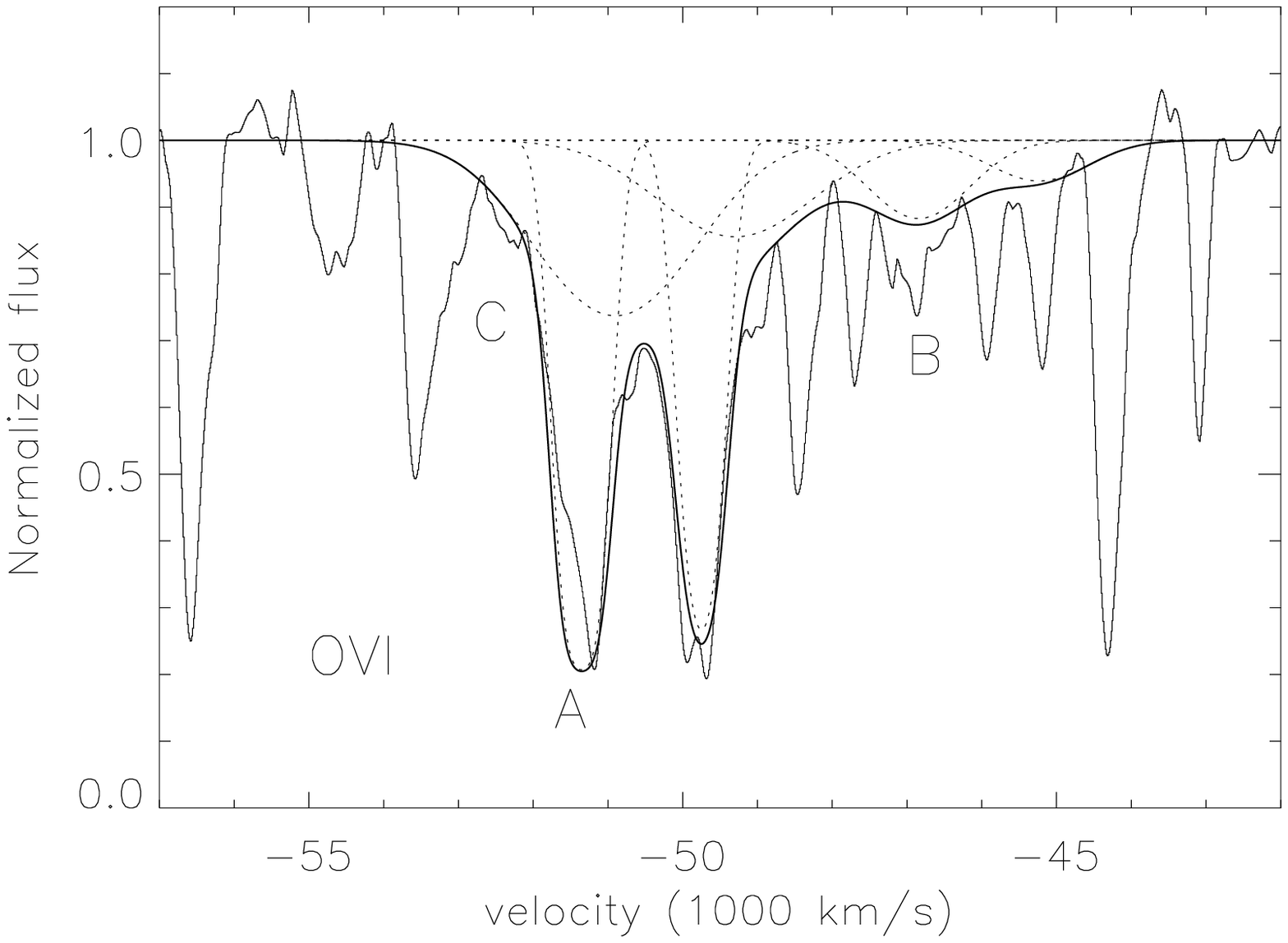}
   \includegraphics[width=9cm]{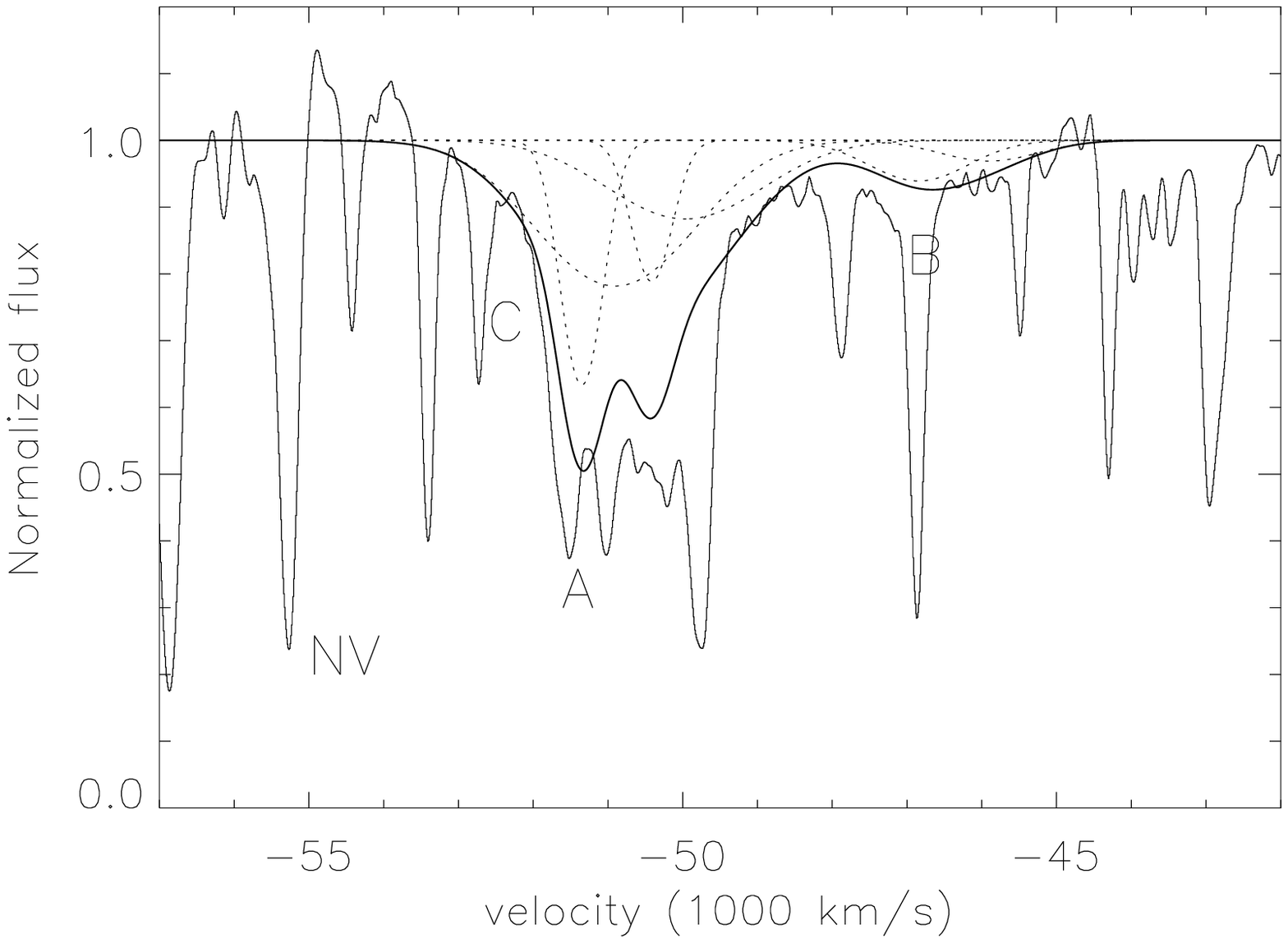}
\caption[Line fitting to \CIV\, in the 1996 Lick spectrum, \OVI\,  and \NV]{Line fitting to \CIV\, in the 1996 Lick spectrum (top), \OVI\, (middle) and \NV\, (bottom). Thick lines represent the results of the combination of the three components of the absorption (A, B, and C - dotted curves). A coverage fraction of $C_f =$ 0.8 was used for every component of every ion. The \CIV\, narrow absorption feature at bluer wavelengths than the \CIV\, Lick absorption is the unrelated system at $z_{abs} \approx$ 1.465.}
\label{fit1}
\end{center}
\end{figure}

Figure \ref{fit1} shows the fit results for \CIV\, \NV, and \OVI\ in the normalized HST and 1996 Lick spectra. The smooth solid curves represent the overall fits, which are combinations of two doublet pairs (labeled A and B) in \CIV\ and three doublets (A, B and C) in \OVI\ and \NV\ (dotted curves). The doublets were forced to have the correct separations and a 2:1 optical depth ratio for the short:long wavelength components. Unrelated absorption from the Ly$\alpha$ forest were masked. Components A and B in \CIV\ are at redshifts $z_{abs} = 1.496$ and 1.534, respectively. These components are also present in \OVI . However, acceptable fits to the \OVI\, profile required i) an additional broad feature (labeled 'C'  in Figure \ref{fit1}), ii) a small shift of 60 km/s in the centroid of component A compared to \CIV\, (see Table \ref{results} below), iii) a smaller FWHM for component A (compared to this component in \CIV ) while the FWHM of component B remained the same, and iv) $C_f=$~0.8 to explain the roughly 1:1 ratio of the \OVI\, doublet intensities in the resolved profile of component A. Based on these results, we adopted $C_f=$~0.8 across the entire \OVI\, mini-BAL and redid the fit to the \CIV\, also assuming $C_f=$~0.8 for consistency\footnote{We also reevaluated our \CIV\ fit to determine if there can be a significant contribution from component C, as in the \OVI\ profile. We found that the measured spectrum in \CIV , e.g., on the blue side of the C component profile at $v\sim -52000$ to $-$53000 \kms , does not permit a significant amount of component C absorption and therefore this component is not included in the final \CIV\ fit.}. This change from $C_f=$~1 in \CIV\, had almost no effect on the quality of the fit or the derived parameter values in the weaker \CIV\, feature. The profile measured in component A of \CIV\, is actually consistent with values of $C_f$ from $\sim$~0.4 to 1.0, which leads to ambiguity in the line optical depths (discussed below). We use the value of $C_f =$~0.8, which is derived from component A of \OVI, for all ions because it is the only value constrained by our data. However, previous studies have found that coverage fractions might be velocity dependent or differ among several ions (\citealt{Barlow97}; \citealt{Hamann97a}; \citealt{Telfer98}; \citealt{Hamann10}), or the absorber(s) can be inhomogeneous (\citealt{deKool02}; \citealt{Arav05}; \citealt{Sabra05}). We address how different $C_f$ values would affect some of our results below and in the Discussion (\S\ref{ionden} and \S\ref{natkin}).

The \NV\, mini-BAL is clearly present in the HST spectrum but there is severe blending with unrelated lines in the Ly$\alpha$ forest. There are also blending problems across the component B absorption in \OVI . The fits to these features are therefore poorly constrained. To overcome these blending difficulties, we fixed the redshift and Doppler $b$ parameter (from which we derive the FWHM value) of component B in \OVI\ and \NV\ to match the values determined from our \CIV\ fit and scale only $\tau_o$ to obtain the best possible fit to the \OVI\ and \NV\ features. Similarly, we forced $z_{abs}$ and $b$ in both components A and C in the \NV\ feature to have the same values as \OVI\ and scaled only $\tau_o$ to obtain the best fit to \NV. Our final best fits to the \CIV\,, \OVI\, and \NV\, mini-BALs assuming $C_f=$~0.8 are shown in Figure \ref{fit1}.

Table \ref{results} provides some of the derived line properties, including the Rest Equivalent Width (REW) measured from the fits to the entire profiles (thick solid lines in Figure \ref{fit1}), and for each profile component A, B and C the fit centroid velocity, $v$, the FWHM, the ionic column density, $N_{ion}$ and the line-center optical depth, $\tau_o$. The last three quantities are listed for fits assuming either $C_f =$~1 or $C_f =$~0.8. We include the measurements based on $C_f = 1$ for comparison. Except for component A of \OVI, values of $N_{ion}$ and $\tau_o$ would increase if $C_f$ is smaller than our estimate of 0.8. If any of those components is saturated at a $C_f < 0.8$, these larger values of $N_{ion}$ and  $\tau_o$ would become lower limits. 
The uncertainties in the line measurements are dominated by the continuum placement and, for OVI and NV, the amount of blending in the Ly\a forest. We estimate 1-$\sigma$ errors in the REW and $N_{ion}$ measurements based on repeated fits that adjust the local continuum height to plausible higher and lower values. These continuum adjustments are guided by the amount of noise and blending in the spectra across the mini-BALs. They do not include changes to the shape of the pseudo-continuum across the OVI and NV lines, which, as noted in \S\ref{cont_norm}, are designed to yield conservatively small values of the line strengths particularly in the wings. Many of the parameters listed for NV and OVI in Table \ref{results} are upper limits because of uncertain amounts of blending in these lines with the Ly$\alpha$ forest. Two exceptions are the values of $N_{ion}$ and $\tau_o$ in OVI component A. With the assumption of $C_f=1$, we derive explicit values for these quantities with the caveat that the fit does not match the apparent 1:1 ratio in the doublet line strengths. A better fit is achieved with $C_f=0.8$, which implies line saturation and minimum values of $N_{ion}$ and $\tau_o$.

Note that the components A, B and C are defined only for convenience in the HST and Lick 1996 data sets. The multi-epoch observations across \CIV\, show that the character of the absorption changes significantly. In particular, new features appear and components A and B lose their identity altogether in the later SDSS and KPNO spectra, as we will discuss later while describing their variability (see sections \S\ref{var_ana} and \S\ref{linevar}). 

\begin{table*}
\begin{minipage}{185mm}
\begin{center}
\caption{Results of the profile fitting of \CIV, \NV, and \OVI. \label{results}}
\scriptsize
\begin{tabular}{c|c|cc|ccc|ccc}
\hline
\hline
 & & & & & C$_f$=1 & & & C$_f$=0.8   &  \\
ion & REW & comp & $v$ & FWHM & N$_{ion}$ & $\tau_o$ & FWHM & N$_{ion}$ & $\tau_o$ \\
& (\AA) & & (km/s) & (km/s) & (10$^{15}$ cm$^{-2}$) & & (km/s) & (10$^{15}$ cm$^{-2}$) & \\
\hline
\CIV &  3.63$^{+0.14}_{-0.16}$ & A & 51260 & 1180 & 0.547$^{+0.011}_{-0.012}$ &  0.18 & 1180 & 0.719$^{+0.016}_{-0.017}$ & 0.25 \\
      &                         &  B & 46820$^a$ & 1580$^a$ & 0.090$^{+0.007}_{-0.010}$ & 0.03 & 1580$^a$ & 0.115$^{+0.008}_{-0.013}$ & 0.04 \\
\NV & $<$5.2               & A & 51320$^b$  & 660$^b$ & $<$ 0.6 & $<$ 0.24 & 680$^b$  & $<$ 0.8 & $<$ 0.33 \\
     &                     & B & 46820$^a$  & 1580$^a$ & $<$ 0.2 & $<$ 0.03 & 1580$^a$ & $<$ 0.2  & $<$ 0.04 \\
     &                     & C & 50860$^c$  & 2500$^c$ & $<$ 1.1 & $<$ 0.11 & 2540$^c$  & $<$ 1.6 & $<$  0.16 \\
\OVI & 7.6$^{+0.4}_{-0.3}$ & A & 51320$^b$  & 720$^b$  & 2.41$^{+0.15}_{-0.15}$ & 0.7 & 780$^b$ & $\gsim$ 8.2 & $\gsim$ 2.8 \\
     &                     & B & 46820$^a$  & 1600$^a$ & $<$ 2.26 & $<$ 0.16 & 1600$^a$ & $<$ 0.72 & $<$ 0.08 \\
    &                     & C & 50860$^c$  & 2510$^c$ & $<$ 0.63 & $<$ 0.07 & 2540$^c$  & $<$ 2.8 & $<$ 0.2 \\
\hline
\end{tabular}
\end{center}
$^a$, $^b$ and $^c$ parameters are locked together (see text).\\
\end{minipage}
\end{table*}

\par
\begin{figure}
\begin{center}
   \includegraphics[width=9cm]{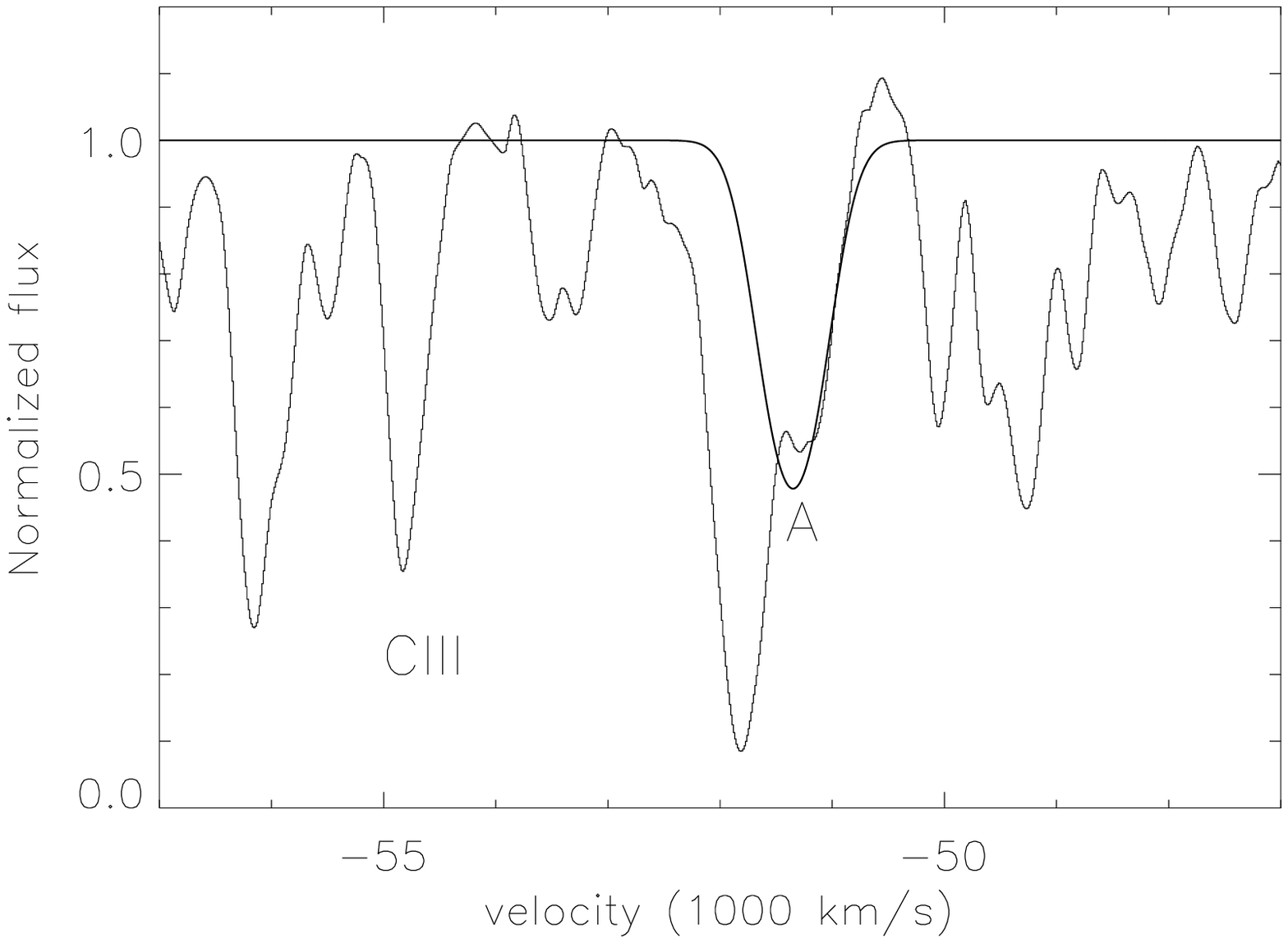}
   \includegraphics[width=9cm]{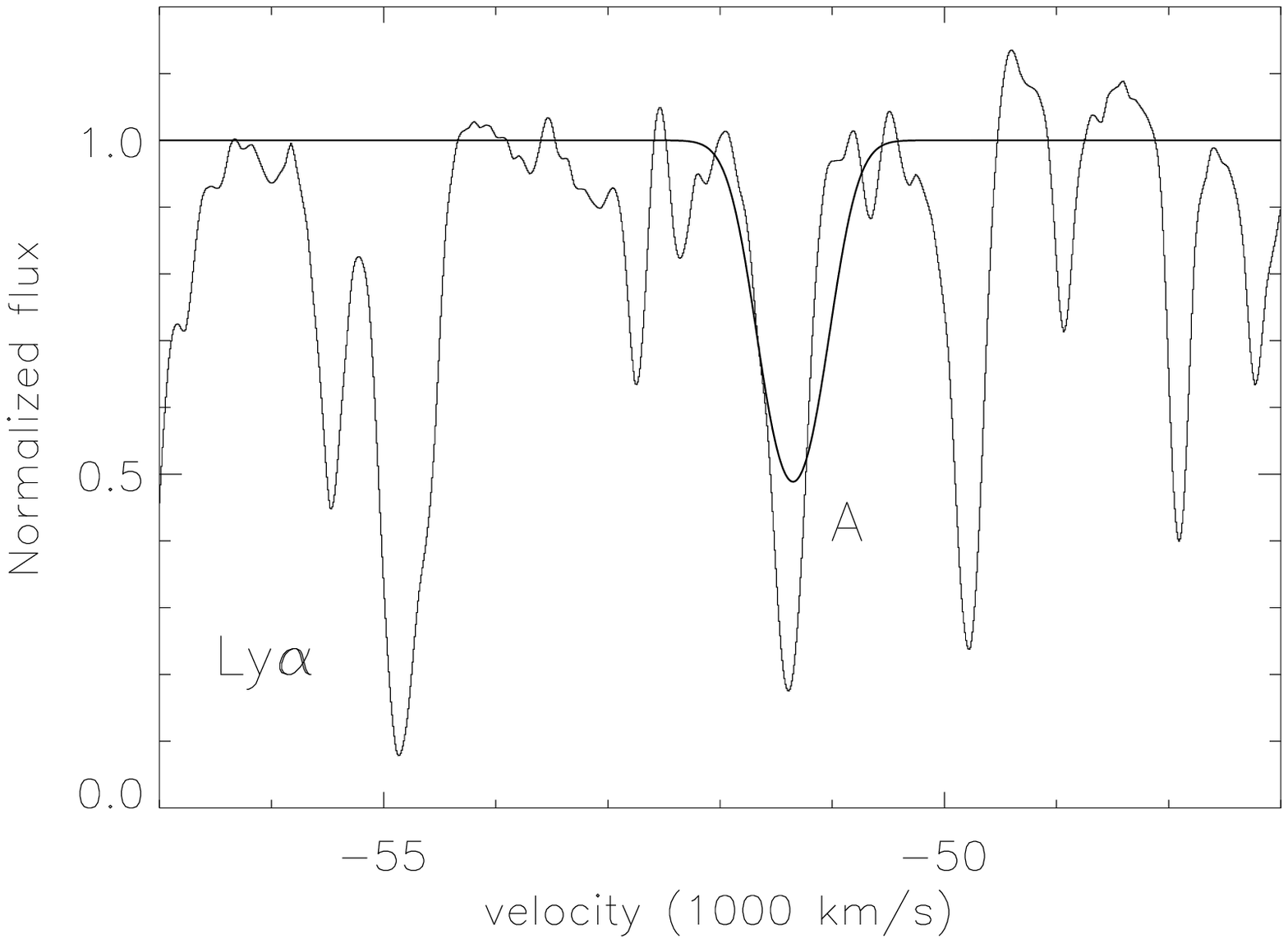}
   \includegraphics[width=9cm]{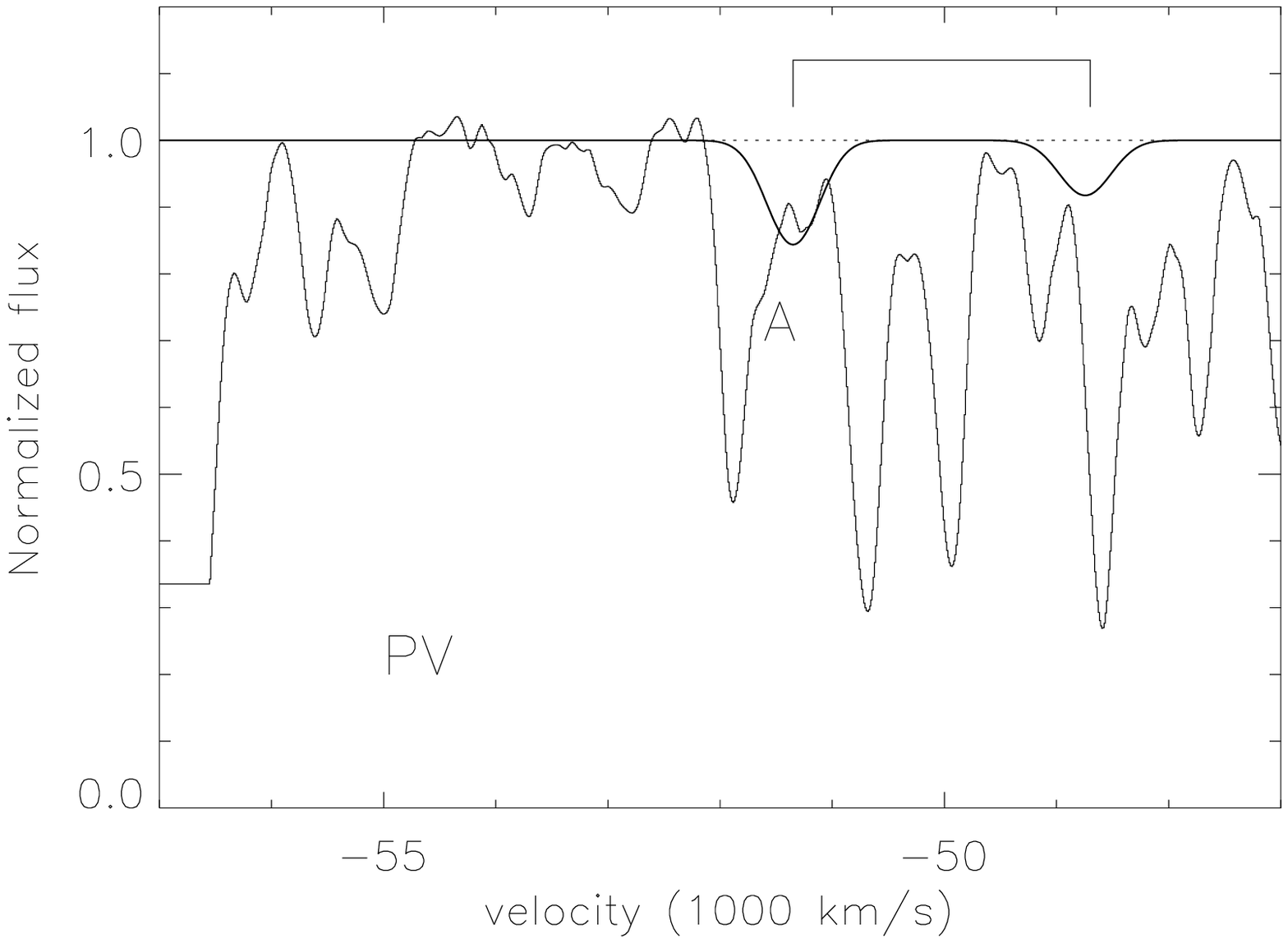}
\caption[Upper limits to \CIII, \NIII, and \PV absorption lines]{Upper limits to \CIII\, (top), Ly\a (middle) and \PV\, (bottom) absorption lines based on $b$ and $z_{abs}$ parameters of component A in \OVI. 
}
\label{NyC}
\end{center}
\end{figure}

 As noted in section \ref{identf}, we do not clearly detect absorption in \CIII, \NIII, \PV, \SiIV, Ly$\alpha$ or Ly$\beta$, at the redshift of the \CIV, \NV\, and \OVI\, absorbers. All of these features lie in the Ly$\alpha$ forest and in every case some absorption is present. Therefore, to place upper limits on the column densities and set limits on the absorber ionization, we simply scale absorption profiles based on the strongest component A at $z_{abs} \approx$~1.496 for all these ions. We use the same $z_{abs}$ and $b$ parameters as in \OVI\, for lines in the HST spectral coverage, and the same parameters as in \CIV\, for lines in the Lick spectra. To place upper limits on these lines, we increase the component A optical depth  until it was obviously contradicted by the data at the wavelengths of interest. Figure \ref{NyC} shows examples of these upper limits and Table \ref{uplim} includes the results obtained by this method. 

We also place an upper limit for the strength of the Lyman limit accompanying the mini-BAL at $z_{abs} \approx$~1.496 (see Fig.\ref{HSTspec}). The Lyman limit for this system is not detected. The sharp absorption edge at 2250 \AA\ (observed) in Fig.\ref{HSTspec} belongs to an unrelated intervening absorption system at $z_{abs} \approx$~1.465 (see Fig.\ref{LICKspec}). We conservatively estimate that the Lyman limit at the redshift of the strongest mini-BAL component (A) has a maximum optical depth of $\tau_{LL}\la 0.2$ (at $\sim$~2276 \AA\ observed, Fig.\ref{HSTspec}). This corresponds to a maximum \HI\, column density of $\sim 3\times 10^{16}$ cm$^{-2}$, which is consistent with the much smaller upper limits we derived from the Lyman lines (Table \ref{uplim}).

\begin{table}
\begin{center}
\caption{Upper limits on lines not detected \label{uplim}}
\scriptsize
\begin{tabular}{c|cc}
\hline
\hline
 & & $C_f$=0.8   \\
ion & REW (\AA) & N$_{ion}$ (10$^{15}$ cm$^{-2}$) \\
\hline
\CIII & $<$ 1.3  & $<$ 0.36 \\
\NIII &  $<$ 0.4 & $<$ 0.55 \\
\PV & $<$ 0.5  &  $<$ 0.09  \\
\SiIV\, & $<$ 1.7 & $<$ 0.17 \\
Ly$\alpha$ & $<$ 1.6 & $<$ 0.5 \\ 
Ly$\beta$ & $<$ 0.8 & $<$ 1.57  \\
\hline
\end{tabular}
\end{center}
\end{table}

\subsection{New Results on the mini-BAL Variability.}
\label{var_ana}

Multi-epoch observations across the \CIV\, feature show variable absorption across a range of  velocities in the outflow. The first published spectrum of PG0935+417, recorded in 1982 (\citealt{Bechtold84}), shows no evidence for the mini-BAL feature. It appeared sometime between the observed epochs of 1982 and 1993. We estimate that the mini-BAL was at least ten times weaker in the 1982 observation compared to its maximum measured strength in 1996 (Fig. 1; \citealt{Narayanan04}). This mini-BAL is therefore another example of an emerging quasar outflow (c.f., \citealt{Hamann08}, \citealt{Leighly09}). It has remained present (albeit variable) up to at least 2008 (Rodriguez Hidalgo et al., in prep.).

\begin{figure}
\begin{center}
\includegraphics[width=8.7cm]{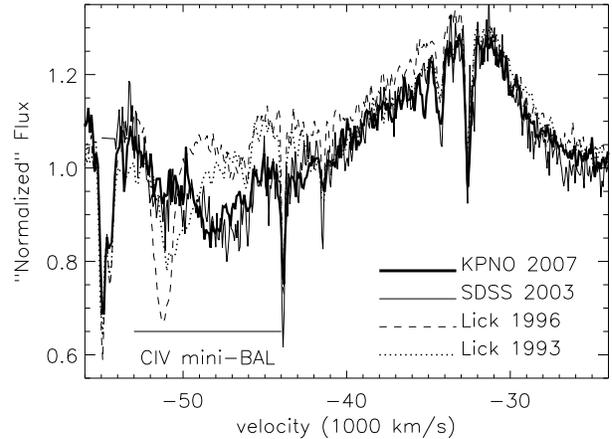}
\caption[Variability in the high velocity \CIV\ $\lambda\lambda$1548,1551 feature over a period of ten years]{Variability in the high velocity \CIV\ $\lambda\lambda$1548,1551 feature ($\sim$3850-3900 \AA\, in observed wavelengths) over a period of fourteen years (4.7 years in the quasar restframe). Emission+continuum around the absorption troughs has been divided by a linear fit to show only variations in absorption, although there may be some remaining emission in the region around of the \CIV\, mini-BAL. The absorption feature at $\sim-50000~\kms$ in the SDSS and KPNO spectra seems weaker than in previous spectra. However, the component outflowing at a slightly lower $v$ ($\sim-47000$ km s$^{-1}$, $\sim$3920 \AA\, in observed wavelength) seems stronger and wider in the two most recent spectra.}
\label{variab}
\end{center}
\end{figure}

Figure \ref{variab} shows two of the spectra included in the \citet{Narayanan04} study (Lick 1993 and 1996) and more recent data from the SDSS (2003) and KPNO (2007). These spectra are plotted after dividing by an estimated linear continuum (excluding all emission and absorption lines) to show better the absorption line changes. We note that in defining this continuum we ignored the entire spectral region that might encompass mini-BAL absorption, from rest wavelengths $\sim$~1290 -- 1350 \AA\ (underlined in Figure \ref{variab}), due to the complexity of the absorber, the wavelength range limits of which are uncertain, and the possible changes of the emission lines in that region. Unfortunately, this implies that both emission and absorption line changes are shown together in Figure \ref{variab}. Nonetheless, inspection of other spectral regions with emission lines (such as \CIV\, and \SiIV) indicates no or small changes. Thus, the variability across the absorption region of interest, marked by the horizontal bar and labeled \CIV\ mini-BAL in Figure \ref{variab}, is likely dominated by the mini-BAL and not by changes in the weak underlying emission lines.

In the 1996 spectrum, we identify distinct mini-BALs in \CIV\, that we labeled A and B (Figure \ref{fit1}). However, these features evolve and lose their identities altogether in subsequent observations. The two most recent spectra show that the variability has continued. The most dramatic changes occurred between 1999 (where the absorption feature is similar to the one in 1996, just slightly weaker, see \citealt{Narayanan04}) and 2003 ($\approx$~1.3 years in the quasar rest-frame). The SDSS 2003 spectrum shows that the absorption profile identified as A in the 1996 spectrum in Figure \ref{fit1} has almost completely disappeared and the absorption previously identified as B has increased in both strength and width. Between 2003 and 2007, the absorption remains quite similar in strength, and there is only a change in the maximum depth, which shifts from $v \sim  -47000$ km s$^{-1}$ in the 2003 spectrum to $v \sim -49000$ km s$^{-1}$ in the 2007 one. Similar shifts in the centroid velocity were reported between the previous Lick observations in \citet{Narayanan04}.

\section{Discussion}
\label{d}

\subsection{Ionization and Total Column Density of the Outflow}
\label{ionden}

\par We parameterize the degree of ionization in the outflow gas based on the ionization parameter, $U$, which is the dimensionless ratio of hydrogen-ionizing photons at the illuminated face of a cloud to the total hydrogen density. We estimate values of $U$ and the total hydrogen column density, $N_H$, by comparing the measured column densities in Tables 2 and 3 (with $C_f=0.8$) to theoretical predictions for a gas that is in photoionization equilibrium with a nominal quasar spectrum and optically thin in the Lyman continuum (\citealt{Hamann10}). The theoretical results from \citet{Hamann10} use the numerical code Cloudy, described by \citet{Ferland98}. We assume that the gas has solar abundances (\citealt{Asplund09}). 

\begin{figure}
\begin{center}
\includegraphics[width=8.7cm]{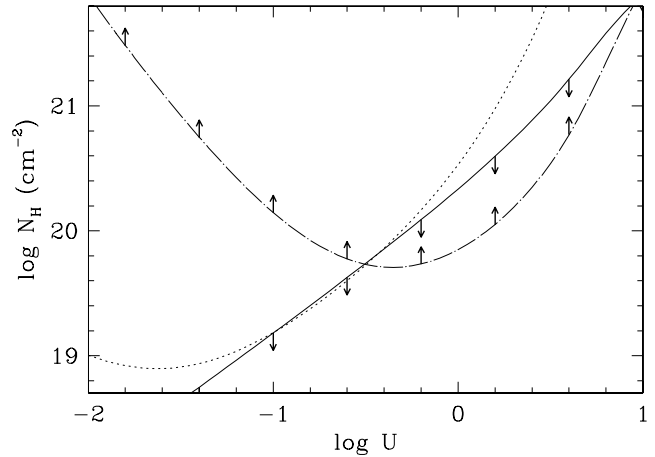}
\caption{Theoretical values of $N_H$ and $U$ in the PG~0935+417 outflow derived from the lower limit on $N$(\OVI) (dash-dot curve), the direct measurement of $N$(\CIV) (dotted curve), and the upper limit on $N$(\HI) determined from Ly$\alpha$ (solid curve). Where the three lines converge, we find our best estimates of these parameters: $\log U \sim -0.5$ and $N_H \sim 5 \times 10^{19}$ cm$^{-2}$.}
\label{NH}
\end{center}
\end{figure}

Figure \ref{NH} shows the theoretical values of $U$ and $N_H$ that are consistent with the measured value of $N_{ion}$ in \CIV\ component A (dotted curve), the lower limit on $N_{ion}$ in \OVI\ component A (dash-dot curve), and the upper limit on \HI\ (solid curve). The arrows on the \OVI\ and \HI\ curves point to regions of the diagram that are consistent with the limits on these ionic column densities. The region between the \OVI\ and \HI\ curves, between roughly $-0.5\lsim \log U\lsim +0.9$, represents the range of parameters that are consistent with {\it both} the lower limit on \OVI\ and the upper limit on \HI . The dotted curve representing the direct measurement of \CIV\ marginally intersects this ``allowed" region of the diagram near $\log U\approx -0.5$. This leads to our best estimates of the ionization, $\log U\approx -0.5$, and total column density, $N_H\approx 5\times 10^{19}$ cm$^{-2}$. We note, however, that this comparison of the \CIV\ results to \OVI\ and \HI\ is uncertain because of several factors. First, the data were not obtained simulateously (\S\ref{obs}) and the profiles differ (\S\ref{fits}). Second, a metallicity different from solar would shift the vertical positions of the \CIV\ and \OVI\ curves in the figure without shifting the \HI\ curve. For example, if we assume a metallicity of twice solar, which might be more appropriate for quasar environments and specifically quasar outflows (e.g., \citealt{Hamann99}; \citealt{Dietrich03}; \citealt{Hamann08}), the \CIV\ and \OVI\ curves would both shift downward in Figure \ref{NH} by 0.3 dex in $\log N_H$. This shift would reduce our best estimate of $N_H$ by roughly a factor of two but it would reduce the allowed range in $\log U$ by only about 0.1 dex, to $\log U\sim -0.6$. Finally, if \CIV\, and/or \HI\, have different values of $C_f$ (see \S\ref{fits}) their curves would shift. For example, if $C_f$(\CIV) were 0.4, the \CIV\, curve in Figure \ref{NH} would shift upwards by 0.4 dex in $\log N_H$. Then, this curve and the \OVI\, curve would intersect around $N_H \sim 6.5\times 10^{19}$ cm$^{-2}$ and $\log U\sim -0.7$. On the contrary, if $C_f$(\CIV) were 1, the shift would occur downwards by a mere 0.12 dex in $\log N_H$, and the intersection of the \CIV\, and \OVI\, curves would happen at a similar $N_H$ than for the case of $C_f=$~0.8, while there would be a slight increase in $\log U$ to $\sim$~$-0.45$.

\subsection{Kinematics and the Nature of the Absorber}
\label{natkin}

The mini-BALs detected in \CIV, \NV\, and \OVI\, span outflow velocities from roughly $-45000$ to $-54000$ \kms. Our measurements of the \NV\, and \OVI\, features confirm the identification of the \CIV\, mini-BAL and provide additional constraints on the nature of this outflow. A comparison of the \CIV\, to \OVI\, and \NV\, lines, in 1996(Lick) and 1994/1995(HST) respectively, also suggests that the amount of absorption at different velocities has an ionization dependence. In particular, the narrow component (A) is narrower in \OVI\, (FWHM~$\sim$~750 km s$^{-1}$) than in \CIV\, (FWHM~$\sim$~1200 km s$^{-1}$) and the \OVI\, and \NV\, absorption features require an additional broad feature (component C with FWHM~$\sim$~2500 km s$^{-1}$) to fit both profiles (\S\ref{fits}). These differences might be affected by the line variabilities since the spectra were taken at different times ($\sim$1.5 months apart in the quasar rest frame). However, we note that the \OVI\, absorption did not change significantly between the two HST observations in 1994 and 1995, and although the \CIV\, line measured did vary from 1993 to 1996, in both cases it had a consistently broader component A and no very broad component C. Finally, all three ions show absorption at lower velocities (component B in Figure \ref{fit1}), which is best matched by an absorption profile with $v \sim -47000$ km s$^{-1}$ ($z_{abs} \sim$ 1.534) and FWHM~$\sim$~1600~$\kms$.

Our analysis is based on our best estimate for the coverage fraction ($C_f =$ 0.8). This value was derived from the mild saturation of component A in \OVI\, (see \S\ref{fits}). The \OVI\, absorption occurs at wavelengths away from the broad emission lines and therefore the absorber must partially cover the quasar continuum source, which at these wavelengths ($\sim$870 \AA\, in the rest frame of the quasar) should have a radius of approximately 0.005 pc if the emission arises from an optically thick, geometrically thin accretion disk (\citealt{Peterson97}). We derive this estimate of the continuum size assuming i) the luminosity is $L \sim$ 6 x 10$^{47}$ erg s$^{-1}$ for PG0935+417 based on the observed flux at 1450 \AA\ (rest-frame) in the absolute flux-calibrated SDSS spectra, a cosmology with $H_o = 71$ \kms\ Mpc, $\Omega_M = 0.27$, $\Omega_{\Lambda}=0.73$, and a bolometric correction factor $L\approx 4.36\lambda L_{\lambda}(1450 {\rm \AA})$ (\citealt{Warner04}; \citealt{Hamann10}), ii) the central black hole mass is $M_{BH} \sim$ 2 x 10$^{10}$ $M_{\odot}$ based on the FWHM of the \CIV\, emission line and $\lambda L_{\lambda}(1450 {\rm \AA})$ (\citealt{Warner03}; \citealt{Vestergaard06}), and iii) the radiative efficiency of the disk is $\eta = 0.1$. As \citet{Vestergaard06} point out, because $M_{BH}$ is obtained by mass scaling relationships it could be off by an order of magnitude, thus it should be taken as a crude estimate. An increase/decrease of one order of magnitude of $M_{BH}$ would result in double/half our estimate of the continuum size.
The value for the accretion efficiency factor of $\eta = 0.1$ is also uncertain but likely to be of the right order of magnitude (\citealt{Peterson97}). Therefore, the observed partial coverage indicates that the characteristic size of outflow absorbing structures that cause this partial covering by \OVI\, in PG0935+417 is $\sim$0.005 pc based on the size of the continuum at the \OVI\, absorbing wavelengths. The absorbing structures could be smaller than this if they have a patchy spatial distribution, such that the characteristic size derived above represents the total projected area of absorbing material across the emitter. If the absorbing material is inhomogeneous and more smoothly distributed across the emission source than patchy clouds, then the 
characteristic size inferred above would represent an approximate area over which the absorber is optically thick in a given line (\citealt{Hamann04}). In any case, partial covering by clouds much larger than the emission source seems unlikely because the clouds would need to have both extremely sharp edges and fortuitous spatial alignments.

\subsection{Variability and Location}
\label{linevar}

Variability helps to confirm the outflowing nature of this absorber. As explained in section \S\ref{var_ana}, the high-velocity absorption profiles are complex and highly variable, such that the absorption components A and B identified circa 1996 in \CIV\ lose their distinct identities in other epochs. In particular, they do not have clear meanings in the spectra beyond 1999. We cannot rule out acceleration/deceleration as the cause of these changes, but since the absorption troughs are complex and they are not shifting as discrete well-defined components (as it was reported in \citealt{Gabel03} and \citealt{Hall07}), we do not interpret the line changes in terms of acceleration or deceleration. Instead, the profile changes appear to be the result of weakening of the absorption at some velocities together with a strengthening at other velocities. The centroid shifts look more like changes in a velocity-dependent $\tau_v$ in a complex absorber rather than real changes in the speed of particular gas components in the outflow. We will present a more detailed analysis of the monitoring of this and other mini-BALs variability in an future paper.

The combined properties  of broad smooth absorption troughs, line variability on time scales $\leq$ 1 yr in the quasar rest frame, and extremely small clouds implied by partial covering of the quasar continuum source strongly suggest that the high velocity absorption forms in an outflow very near the quasar (e.g., \citealt{Hamann97a} and \citealt{Hamann10}). The variability times, in particular, help to provide direct constraints on the absorber location (\citealt{Hamann97a}). If the flow is photoionized and close to ionization equilibrium, and the line changes are caused by small changes in the ionizing flux, then the recombination time sets an approximate limit on the outflow gas density: $\tau_{rec} \sim \frac{1}{\alpha_r n_e}$, where $\alpha_r$ is the recombination rate coefficient for \CIV\, in this case, and $n_e$ is the electron density. Using the shortest variability time that we measure ($\sim$1 yr, in the quasar rest frame) as an upper limit for the recombination time, with a value for $\alpha_r$ of $2.8 \times 10^{-12}$ cm$^{-3}$ s$^{-1}$ (\citealt{Arnaud85}) for \CIV\, (assuming this is the predominant \C\, ion which is likely at $\log U \gsim -0.5$), we obtain a minimum electron density $n_H \gsim 1.1 \times 10^{4}$ cm$^{-3}$. This, in turn, yields a maximum distance between the absorber and the quasar emission source because the ionization parameter scales like $U \propto \frac{L}{n_H\, r^2}$ (\citealt{Narayanan04}; \citealt{Hamann10b}), where $L$ is the quasar luminosity, $n_H \sim n_e$ is the absorber density, and $r$ is the radial distance. Thus, we derive a maximum distance of $r \lsim$ 1.2 kpc using $\log U \gsim -0.5$ and $n_H \gsim 1.1 \times 10^{4}$ cm$^{-3}$. 

If, on the other hand, the mini-BAL changes are caused by clouds moving across our lines of sight to the continuum source, as proposed recently by \citet{Hamann08} and \citet{Hamann10}, then the transverse speeds (perpendicular to our sightlines) must be large enough to cross a significant portion of the far-UV emission source in the variability time.
In a scenario where the absorber has a simple shape and is homogeneous, we can expect a significant variation of the absorption profile when the absorber has crossed half of the path in front of the continuum source. The smallest observed variability time, $\sim$1 yr, and the continuum region diameter at the wavelengths of the high velocity absorber\footnote{Obtained also from eqn. 3.18 and 3.21 in \citet{Peterson97}, similarly to what we did in \S\ref{natkin} for the continuum around the \OVI\, absorption.}, $D_{1300} \sim$0.02 pc, require transverse velocities $v_{tr} \sim$~10000 km s$^{-1}$ to cross half of the continuum source. Assuming this transverse speed equals the Keplerian speed for the absorber's location relative to the center of mass, it can help constrain the location of the outflow and it would locate the absorber at a radius of $r \lsim$~0.9 pc. Thus, this would place the outflow outside or in the vicinity of the \CIV\, BEL region radius ($R_{\rm BELR} \sim$~0.5 pc; \citealt{Hamann10b}), as current models seem to locate it (\citealt{Proga00}; \citealt{Kurosawa09}).

While both scenarios are plausible, we would like to note that changes in the ionizing flux might have difficulties to provoke the range of complex profile changes we observe. Continued monitoring can test these scenarios  because simple ionization changes should lead to repeating patterns of variability, as those observed in \citet{Misawa07b}.  

\subsection{Comparisons to other quasar outflows}
\label{relother}

Mini-BALs are almost as common in quasar spectra as BALs (\citealt{RodriguezHidalgo09}; Rodriguez Hidalgo et al., in prep.). Mini-BALs might be produced in the same general outflow as BALs, but viewed along different lines of sight to the continuum source. For example, they might appear along sightlines farther above the accretion disk, that miss the main part of the BAL flow nearer the disk plane (\citealt{Ganguly01}). \citet{Hamann08} proposed that instabilities and filamentary absorbing structures residing near this ragged edge of the BAL flow could provide a natural explanation for the dramatic variability observed in some high-velocity mini-BALs and detached BALs. Perhaps this picture applies to the mini-BAL outflow in PG0935+417.

The most puzzling aspect of this outflow is its extreme high speed, nominally around $-50000 \kms$. While mini-BAL outflows overall are fairly common, observed speeds $|v| \geq 50000$ km s$^{-1}$ are extremely rare. Mini-BALs with speeds $|v| > 25000$ km s$^{-1}$ are found in only $\sim2.6\pm0.5$\% of optically selected quasars, compared to $\sim9.9\pm0.7$\% having mini-BALs at $|v| < 25000$ \kms (\citealt{RodriguezHidalgo09}; Rodriguez Hidalgo et al., in prep). Nonetheless, there does not appear to be anything unusual about PG~0935+417 that would produce an unusually fast flow. It does have a high luminosity, but the Eddington ratio is very modest, $L / L_{Edd} \sim$ 0.2 (based on our value of $M_{BH} \sim$ 2 x 10$^{10}$ $M_{\odot}$, \S4.2). It is also radio-quiet, it has a standard blue spectral shape and ordinary looking broad emission lines for a luminous quasar. 

One interesting property of PG~0935+417 is that it appears to have much less X-ray absorption than typical BAL quasars. In particular, Rodriguez Hidalgo (in prep.) shows that PG~0935+417 has a two point spectral index between 2500 \AA\ and 2 keV in the rest frame of $\alpha_{ox}\sim$~-1.71 based on a measurement with the {\it X-ray Multi-Mirror Mission (XMM)} in 2007. This value of $\alpha_{ox}$ is typical of quasars without BALs (\citealt{Steffen06}; \citealt{Chartas09}; \citealt{Gibson10}), but it is much less steep than BAL quasars where the X-ray flux is severely suppressed by absorption (\citealt{Gallagher02, Gallagher06}). Explicit fits to the X-ray spectrum of PG~0935+417 indicate that the amount of X-ray absorption is limited to $N_H\lsim 2\times 10^{22}$ cm$^{-2}$ at 90\% confidence or $N_H\lsim 4\times 10^{22}$ cm$^{-2}$ at 99\% (for an equivalent neutral absorber). This upper limit on the X-ray absorption in PG~0935+417 is at least several times smaller than the lower limits derived typically for BAL quasars (see also \citealt{Mathur00}). 

These results from the X-ray observations have important consequences for the structure and acceleration of quasar outflows. One common view is that the outflow gas is accelerated by radiation pressure behind a thick radiative shield that blocks much of the X-ray and far-UV flux from the central quasar (\citealt{Murray95, Murray97}). If the shield is not present or not optically thick enough, the outflow gas can become too ionized and too transparent to be driven radiatively to high speeds. We might expect, therefore, that an extreme high-velocity flow like the one in PG~0935+417 should be accompanied by X-ray absorption that is at least as strong as BAL quasars (where the typical flow speeds are 2 to 3 times smaller than PG~0935+417). Indeed, \cite{Gallagher06} report a tentative trend among BAL quasars for stronger X-ray absorption in sources with higher maximum flow velocities. However, the observations of PG~0935+417 demonstrate that extremely high outflow speeds can be achieved even if the amount of X-ray absorption is at most moderate (or perhaps absent altogether). 

This situation is similar to other mini-BAL outflows, which generally have weak or absent X-ray absorption (\citealt{Misawa08}; \citealt{Chartas09}; \citealt{Gibson10}), but the extreme speeds in PG~0935+417 present a more serious challenge to our theoretical understanding of how these outflows work. \citet{Chartas09} favor a scenario in which the mini-BAL gas follows curved trajectories, such that it was accelerated in a thickly shielded environment but now appears along sightlines to the quasar that are relatively free of X-ray absorption. However, \citet{Hamann10} noted that the observed ionizations in mini-BALs are just like typical BALs, even though mini-BALs do not reside behind a strong radiative shield. Therefore, radiative shielding in the X-rays and far-UV is not essential to moderate the outflow ionization and, evidently, it is also not essential for the radiative acceleration of quasar outflows to high speeds.

Finally, it is interesting that the \CIV\ mini-BAL remained at approximately the same velocity (with no apparent acceleration or deceleration, \S4.3) throughout the observation period from at least 1993 to 2007, corresponding to $\gsim$4.7 years in the quasar frame. At a nominal speed of $-50000$ \kms , the distance traveled by the outflow gas during this time is roughly 0.24 pc. This is a significant fraction of the maximum radial distance of the flow, $r \lsim 1$ pc, derived above for clouds crossing our lines of sight (\S4.3). If the observed mini-BALs do form in blobs or filaments created via instabilities in the inner flow (\citealt{Hamann08}; \citealt{Hamann10}), perhaps resembling the flow structures seen in some numerical simulations (\citealt{Proga00}; \citealt{Kurosawa09}), then this dramatic mini-BAL variability is expected as these structures dissipate or evolve significantly during observing periods that approach the flow crossing time of $t_f\sim r/v \lsim 18$ yrs. Continued monitoring of the mini-BALs in PG~0935+417 should provide valuable constraints on the location and basic nature of these unusual outflow structures. 

\section{Summary and Conclusions}

Energetic outflows are common phenomena in AGN that could help us understand the mechanisms that power the central engines and, perhaps, provide feedback to the host galaxy's evolution. We present new measurements and analyses of outflow absorption lines in spectra of the quasar PG0935+417 ($z_{em}\approx$ 1.966). This quasar shows an extreme high-velocity \CIV\, mini-Broad Absorption Line (mini-BAL) outflowing at $\sim -50000$ km s$^{-1}$. We detect, using Lick observatory, Hubble Space Telescope, and Sloan Digital Sky Survey spectra, absorption in \CIVdbl\ (already reported in \citealt{Hamann97b} and  \citealt{Narayanan04}) and, for the first time in this outflow, in \NVdbl\, and \OVIdbl\, in this mini-BAL system. We also present new results on the outflow variability. 

\begin{itemize}

\item {\it Ionization and Total Column Density}. Our measurements and limits on $N$(\OVI), $N$(\CIV) and $N$(Ly$\alpha$) indicate that the outflow has ionization $\log U \sim -0.5$ and total column density of $N_H \sim 5 \times 10^{19} 
$ cm$^{-2}$ (see Figure \ref{NH}), within some uncertainties (\S\ref{ionden}). The absence of lower ionization lines (where we place also upper limits) confirms that the flow is highly ionized.

\item {\it Kinematics and Absorber Structure}. The \CIV, \NV, and \OVI\, absorption profiles are detected across a range of velocities from $-45000$ to $-54000$ \kms. Table \ref{results} shows our measurements of REW, velocity, FWHM, $N_{ion}$ and $\tau_o$ for line components we identified in the particular epoch of the 1996 Lick and 1994/1995 HST observations. The resolved component A of \OVI\, indicates that the lines are moderately saturated with the absorber covering just $\sim$80\% of the background continuum source. Since the \OVI\, absorption occurs at wavelengths away from the broad emission lines, this partial covering result implies that the absorber must have a characteristic size smaller than the quasar continuum source, which we estimate to have a radius of roughly 0.005 pc. 

\item {\it Variability and Location}. The high-velocity mini-BAL profiles are complex and highly variable, from the time the outflow emerged (between 1982 and 1993) to the most recent observation (2007). The shortest observed variability time is approximately 1 year in the quasar rest-frame. If the variability is caused by changes in the ionization, this timescale results in a location constraint of $r \lsim$~1.2 kpc from the continuum source and a density of $n_H \gsim 1.1 \times 10^{4}$ cm$^{-3}$. However, if the variability is cauqsed by motions of absorbing clouds  across the line of sight, the maximum radial distance is only $r\lsim 0.9$ pc. 

\item {\it The Nature of the Outflow}. Extreme high-velocity outflows like the one in PG~0935+417 present unique challenges to our understanding of the physics behind  the quasar outflow phenomenon. Outflows at these speeds are rarely detected in quasar spectra, but they might be common in quasars if the gas subtends just a small fraction of the sky as seen from the central continuum source. PG~0935+417 does not appear to have any unusual properties (compared to other quasars) that might explain the rare occurrence of an extreme outflow in this quasar. The absence of strong X-ray absorption in PG~0935+417 is in marked contrast to BAL quasars but typical of other quasars with mini-BAL or NAL outflows. It suggests that a change from the standard paradigm of quasar outflows is needed, namely that strong radiative shielding in the X-rays and far-UV may not be important for the acceleration of the gas to high speeds.

\end{itemize}

\section*{Acknowledgments}

P.R.H. would like to thank Mike Crenshaw for discussions and comments to the manuscript, and the Alumni Fellowship at UF for partial financial support for this project. P.R.H. and F.H. would like to thank George Chartas for discussions on the XMM data. We thank an anonymous referee for valuable comments. P.R.H and F.H. acknowledge support from Chandra award TM9-0005X and from XMM-Newton program number 50462. P.B.H. is supported by NSERC. Funding for the SDSS and SDSS-II has been provided by the Alfred P. Sloan Foundation, the Participating Institutions, the National Science Foundation, the U.S. Department of Energy, the National Aeronautics and Space Administration, the Japanese Monbukagakusho, the Max Planck Society, and the Higher Education Funding Council for England. The SDSS Web Site is http://www.sdss.org/.

\bibliographystyle{mn2e}
\bibliography{PG0935_astroph}

\begin{thebibliography}{81}
\expandafter\ifx\csname natexlab\endcsname\relax\def\natexlab#1{#1}\fi

\bibitem[{{Adelman-McCarthy} \& {et al.}(2008)}]{Adelman-McCarthy08}
{Adelman-McCarthy} J.~K., {et al.}, 2008, \apjs, 175, 297

\bibitem[{{Aldcroft} {et~al.}(1994){Aldcroft}, {Bechtold}, \&
  {Elvis}}]{Aldcroft94}
{Aldcroft} T.~L., {Bechtold} J., {Elvis} M., 1994, \apjs, 93, 1

\bibitem[{{Arav} {et~al.}(2001){Arav}, {de Kool}, {Korista}, {Crenshaw}, {van
  Breugel}, {Brotherton}, {Green}, {Pettini}, \& {et al.}}]{Arav01}
{Arav} N., {de Kool} M., {Korista} K.~T., {Crenshaw} D.~M., {van Breugel} W.,
  {Brotherton} M., {Green} R.~F., {Pettini} M., {et al.}, 2001, \apj, 561, 118

\bibitem[{{Arav} {et~al.}(2005){Arav}, {Kaastra}, {Kriss}, {Korista}, {Gabel},
  \& {Proga}}]{Arav05}
{Arav} N., {Kaastra} J., {Kriss} G.~A., {Korista} K.~T., {Gabel} J., {Proga}
  D., 2005, \apj, 620, 665

\bibitem[{{Arav} {et~al.}(1994){Arav}, {Li}, \& {Begelman}}]{Arav94}
{Arav} N., {Li} Z.-Y., {Begelman} M.~C., 1994, \apj, 432, 62

\bibitem[{{Arnaud} \& {Rothenflug}(1985)}]{Arnaud85}
{Arnaud} M., {Rothenflug} R., 1985, \aaps, 60, 425

\bibitem[{{Asplund} {et~al.}(2009){Asplund}, {Grevesse}, {Sauval}, \&
  {Scott}}]{Asplund09}
{Asplund} M., {Grevesse} N., {Sauval} A.~J., {Scott} P., 2009, \araa, 47, 481

\bibitem[{{Barlow} \& {Sargent}(1997)}]{Barlow97}
{Barlow} T.~A., {Sargent} W.~L.~W., 1997, \aj, 113, 136

\bibitem[{{Bechtold} {et~al.}(1984){Bechtold}, {Green}, {Weymann}, {Schmidt},
  {Estabrook}, {Sherman}, {Wahlquist}, \& {Heckman}}]{Bechtold84}
{Bechtold} J., {Green} R.~F., {Weymann} R.~J., {Schmidt} M., {Estabrook} F.~B.,
  {Sherman} R.~D., {Wahlquist} H.~D., {Heckman} T.~M., 1984, \apj, 281, 76

\bibitem[{{Chartas} {et~al.}(2002){Chartas}, {Brandt}, {Gallagher}, \&
  {Garmire}}]{Chartas02}
{Chartas} G., {Brandt} W.~N., {Gallagher} S.~C., {Garmire} G.~P., 2002, \apj,
  579, 169

\bibitem[{{Chartas} {et~al.}(2009){Chartas}, {Charlton}, {Eracleous},
  {Giustini}, {Hidalgo}, {Ganguly}, {Hamann}, {Misawa}, \&
  {Tytler}}]{Chartas09}
{Chartas} G., {Charlton} J., {Eracleous} M., {Giustini} M., {Hidalgo} P.~R.,
  {Ganguly} R., {Hamann} F., {Misawa} T., {Tytler} D., 2009, New Astronomy
  Review, 53, 128

\bibitem[{{Churchill} {et~al.}(1999){Churchill}, {Schneider}, {Schmidt}, \&
  {Gunn}}]{Churchill99}
{Churchill} C.~W., {Schneider} D.~P., {Schmidt} M., {Gunn} J.~E., 1999, \aj,
  117, 2573

\bibitem[{{Crenshaw} {et~al.}(1999){Crenshaw}, {Kraemer}, {Boggess}, {Maran},
  {Mushotzky}, \& {Wu}}]{Crenshaw99}
{Crenshaw} D.~M., {Kraemer} S.~B., {Boggess} A., {Maran} S.~P., {Mushotzky}
  R.~F., {Wu} C.-C., 1999, \apj, 516, 750

\bibitem[{{de Kool} \& {Begelman}(1995)}]{deKool95}
{de Kool} M., {Begelman} M.~C., 1995, \apj, 455, 448

\bibitem[{{de Kool} {et~al.}(2002){de Kool}, {Korista}, \& {Arav}}]{deKool02}
{de Kool} M., {Korista} K.~T., {Arav} N., 2002, \apj, 580, 54

\bibitem[{{Di Matteo} {et~al.}(2005){Di Matteo}, {Springel}, \&
  {Hernquist}}]{DiMatteo05}
{Di Matteo} T., {Springel} V., {Hernquist} L., 2005, \nat, 433, 604

\bibitem[{{Dietrich} {et~al.}(2003){Dietrich}, {Hamann}, {Shields},
  {Constantin}, {Heidt}, {J{\"a}ger}, {Vestergaard}, \& {Wagner}}]{Dietrich03}
{Dietrich} M., {Hamann} F., {Shields} J.~C., {Constantin} A., {Heidt} J.,
  {J{\"a}ger} K., {Vestergaard} M., {Wagner} S.~J., 2003, \apj, 589, 722

\bibitem[{{Dunn} {et~al.}(2008){Dunn}, {Crenshaw}, {Kraemer}, \&
  {Trippe}}]{Dunn08}
{Dunn} J.~P., {Crenshaw} D.~M., {Kraemer} S.~B., {Trippe} M.~L., 2008, \aj,
  136, 1201

\bibitem[{{Everett}(2005)}]{Everett05}
{Everett} J.~E., 2005, \apj, 631, 689

\bibitem[{{Ferland} {et~al.}(1998){Ferland}, {Korista}, {Verner}, {Ferguson},
  {Kingdon}, \& {Verner}}]{Ferland98}
{Ferland} G.~J., {Korista} K.~T., {Verner} D.~A., {Ferguson} J.~W., {Kingdon}
  J.~B., {Verner} E.~M., 1998, \pasp, 110, 761

\bibitem[{{Foltz} {et~al.}(1986){Foltz}, {Weymann}, {Peterson}, {Sun},
  {Malkan}, \& {Chaffee}}]{Foltz86}
{Foltz} C.~B., {Weymann} R.~J., {Peterson} B.~M., {Sun} L., {Malkan} M.~A.,
  {Chaffee} Jr. F.~H., 1986, \apj, 307, 504

\bibitem[{{Gabel} {et~al.}(2003){Gabel}, {Crenshaw}, {Kraemer}, {Brandt},
  {George}, {Hamann}, {Kaiser}, {Kaspi}, {Kriss}, {Mathur}, {Mushotzky},
  {Nandra}, {Netzer}, {Peterson}, {Shields}, {Turner}, \& {Zheng}}]{Gabel03}
{Gabel} J.~R., {Crenshaw} D.~M., {Kraemer} S.~B., {Brandt} W.~N., {George}
  I.~M., {Hamann} F.~W., {Kaiser} M.~E., {Kaspi} S., {Kriss} G.~A., {Mathur}
  S., {Mushotzky} R.~F., {Nandra} K., {Netzer} H., {Peterson} B.~M., {Shields}
  J.~C., {Turner} T.~J., {Zheng} W., 2003, \apj, 583, 178

\bibitem[{{Gallagher} {et~al.}(2002){Gallagher}, {Brandt}, {Chartas}, \&
  {Garmire}}]{Gallagher02}
{Gallagher} S.~C., {Brandt} W.~N., {Chartas} G., {Garmire} G.~P., 2002, \apj,
  567, 37

\bibitem[{{Gallagher} {et~al.}(2006){Gallagher}, {Brandt}, {Chartas},
  {Priddey}, {Garmire}, \& {Sambruna}}]{Gallagher06}
{Gallagher} S.~C., {Brandt} W.~N., {Chartas} G., {Priddey} R., {Garmire} G.~P.,
  {Sambruna} R.~M., 2006, \apj, 644, 709

\bibitem[{{Ganguly} {et~al.}(2001){Ganguly}, {Bond}, {Charlton}, {Eracleous},
  {Brandt}, \& {Churchill}}]{Ganguly01}
{Ganguly} R., {Bond} N.~A., {Charlton} J.~C., {Eracleous} M., {Brandt} W.~N.,
  {Churchill} C.~W., 2001, \apj, 549, 133

\bibitem[{{Ganguly} \& {Brotherton}(2008)}]{Ganguly08}
{Ganguly} R., {Brotherton} M.~S., 2008, \apj, 672, 102

\bibitem[{{Gebhardt} {et~al.}(2000){Gebhardt}, {Bender}, {Bower}, {Dressler},
  {Faber}, {Filippenko}, {Green}, {Grillmair}, {Ho}, {Kormendy}, {Lauer},
  {Magorrian}, {Pinkney}, {Richstone}, \& {Tremaine}}]{Gebhardt00}
{Gebhardt} K., {Bender} R., {Bower} G., {Dressler} A., {Faber} S.~M.,
  {Filippenko} A.~V., {Green} R., {Grillmair} C., {Ho} L.~C., {Kormendy} J.,
  {Lauer} T.~R., {Magorrian} J., {Pinkney} J., {Richstone} D., {Tremaine} S.,
  2000, \apjl, 539, L13

\bibitem[{{Gibson} {et~al.}(2010){Gibson}, {Brandt}, {Gallagher}, {Hewett}, \&
  {Schneider}}]{Gibson10}
{Gibson} R.~R., {Brandt} W.~N., {Gallagher} S.~C., {Hewett} P.~C., {Schneider}
  D.~P., 2010, \apj, 713, 220

\bibitem[{{Hall} {et~al.}(2007){Hall}, {Sadavoy}, {Hutsemekers}, {Everett}, \&
  {Rafiee}}]{Hall07}
{Hall} P.~B., {Sadavoy} S.~I., {Hutsemekers} D., {Everett} J.~E., {Rafiee} A.,
  2007, \apj, 665, 174

\bibitem[{{Hamann}(1998)}]{Hamann98}
{Hamann} F., 1998, \apj, 500, 798

\bibitem[{{Hamann}(2010)}]{Hamann10c}
---, 2010, \mnras, (in prep.)

\bibitem[{{Hamann} {et~al.}(1997{\natexlab{a}}){Hamann}, {Barlow}, {Cohen},
  {Junkkarinen}, \& {Burbidge}}]{Hamann97b}
{Hamann} F., {Barlow} T., {Cohen} R.~D., {Junkkarinen} V., {Burbidge} E.~M.,
  1997{\natexlab{a}}, in Astronomical Society of the Pacific Conference Series,
  Vol. 128, Mass Ejection from Active Galactic Nuclei, {Arav} N., {Shlosman}
  I., {Weymann} R.~J., eds., pp. 19--+

\bibitem[{{Hamann} {et~al.}(1997{\natexlab{b}}){Hamann}, {Barlow},
  {Junkkarinen}, \& {Burbidge}}]{Hamann97a}
{Hamann} F., {Barlow} T.~A., {Junkkarinen} V., {Burbidge} E.~M.,
  1997{\natexlab{b}}, \apj, 478, 80

\bibitem[{{Hamann} \& {Ferland}(1999)}]{Hamann99}
{Hamann} F., {Ferland} G., 1999, \araa, 37, 487

\bibitem[{{Hamann} {et~al.}(2010){Hamann}, {Kanekar}, {Prochaska}, {Murphy},
  {Milutinovic}, {Ellison}, \& W.}]{Hamann10}
{Hamann} F., {Kanekar} N., {Prochaska} J.~K., {Murphy} M.~T., {Milutinovic} N.,
  {Ellison} S., W. U., 2010, \mnras, (submitted)

\bibitem[{{Hamann} {et~al.}(2008){Hamann}, {Kaplan}, {Hidalgo}, {Prochaska}, \&
  {Herbert-Fort}}]{Hamann08}
{Hamann} F., {Kaplan} K.~F., {Hidalgo} P.~R., {Prochaska} J.~X., {Herbert-Fort}
  S., 2008, \mnras, 391, L39

\bibitem[{{Hamann} {et~al.}(2002){Hamann}, {Korista}, {Ferland}, {Warner}, \&
  {Baldwin}}]{Hamann02}
{Hamann} F., {Korista} K.~T., {Ferland} G.~J., {Warner} C., {Baldwin} J., 2002,
  \apj, 564, 592

\bibitem[{{Hamann} \& {Sabra}(2004)}]{Hamann04}
{Hamann} F., {Sabra} B., 2004, in Astronomical Society of the Pacific
  Conference Series, Vol. 311, AGN Physics with the Sloan Digital Sky Survey,
  {Richards} G.~T., {Hall} P.~B., eds., pp. 203--+

\bibitem[{{Hamann} \& {Simon}(2010)}]{Hamann10b}
{Hamann} F., {Simon} L.~E., 2010, \mnras, (submitted)

\bibitem[{{Hewitt} \& {Burbidge}(1993)}]{Hewitt93}
{Hewitt} A., {Burbidge} G., 1993, \apjs, 87, 451

\bibitem[{{Jannuzi}(2010)}]{Januzzi10}
{Jannuzi} B.~T., 2010, \apj, (in prep.)

\bibitem[{{Jannuzi} {et~al.}(1996){Jannuzi}, {Hartig}, {Kirhakos}, {Sargent},
  {Turnshek}, {Weymann}, {Bahcall}, {Bergeron}, {Boksenberg}, {Savage},
  {Schneider}, \& {Wolfe}}]{Januzzi96}
{Jannuzi} B.~T., {Hartig} G.~F., {Kirhakos} S., {Sargent} W.~L.~W., {Turnshek}
  D.~A., {Weymann} R.~J., {Bahcall} J.~N., {Bergeron} J., {Boksenberg} A.,
  {Savage} B.~D., {Schneider} D.~P., {Wolfe} A.~M., 1996, \apjl, 470, L11+

\bibitem[{{Kurosawa} \& {Proga}(2009)}]{Kurosawa09}
{Kurosawa} R., {Proga} D., 2009, \apj, 693, 1929

\bibitem[{{Lanzetta} {et~al.}(1995){Lanzetta}, {Wolfe}, \&
  {Turnshek}}]{Lanzetta95}
{Lanzetta} K.~M., {Wolfe} A.~M., {Turnshek} D.~A., 1995, \apj, 440, 435

\bibitem[{{Laor} \& {Brandt}(2002)}]{Laor02}
{Laor} A., {Brandt} W.~N., 2002, \apj, 569, 641

\bibitem[{{Leighly} {et~al.}(2009){Leighly}, {Hamann}, {Casebeer}, \&
  {Grupe}}]{Leighly09}
{Leighly} K.~M., {Hamann} F., {Casebeer} D.~A., {Grupe} D., 2009, \apj, 701,
  176

\bibitem[{{Ma}(2002)}]{Ma02}
{Ma} F., 2002, \mnras, 335, L99

\bibitem[{{Mathur} {et~al.}(2000){Mathur}, {Green}, {Arav}, {Brotherton},
  {Crenshaw}, {deKool}, {Elvis}, {Goodrich}, {Hamann}, {Hines}, {Kashyap},
  {Korista}, {Peterson}, {Shields}, {Shlosman}, {van Breugel}, \&
  {Voit}}]{Mathur00}
{Mathur} S., {Green} P.~J., {Arav} N., {Brotherton} M., {Crenshaw} M., {deKool}
  M., {Elvis} M., {Goodrich} R.~W., {Hamann} F., {Hines} D.~C., {Kashyap} V.,
  {Korista} K., {Peterson} B.~M., {Shields} J.~C., {Shlosman} I., {van Breugel}
  W., {Voit} M., 2000, \apjl, 533, L79

\bibitem[{{Merritt} \& {Ferrarese}(2001)}]{Merritt01}
{Merritt} D., {Ferrarese} L., 2001, \apj, 547, 140

\bibitem[{{Misawa} {et~al.}(2007){Misawa}, {Eracleous}, {Charlton}, \&
  {Kashikawa}}]{Misawa07b}
{Misawa} T., {Eracleous} M., {Charlton} J.~C., {Kashikawa} N., 2007, \apj, 660,
  152

\bibitem[{{Misawa} {et~al.}(2008){Misawa}, {Eracleous}, {Chartas}, \&
  {Charlton}}]{Misawa08}
{Misawa} T., {Eracleous} M., {Chartas} G., {Charlton} J.~C., 2008, \apj, 677,
  863

\bibitem[{{Murray} \& {Chiang}(1997)}]{Murray97}
{Murray} N., {Chiang} J., 1997, \apj, 474, 91

\bibitem[{{Murray} {et~al.}(1995){Murray}, {Chiang}, {Grossman}, \&
  {Voit}}]{Murray95}
{Murray} N., {Chiang} J., {Grossman} S.~A., {Voit} G.~M., 1995, \apj, 451, 498

\bibitem[{{Narayanan} {et~al.}(2004){Narayanan}, {Hamann}, {Barlow},
  {Burbidge}, {Cohen}, {Junkkarinen}, \& {Lyons}}]{Narayanan04}
{Narayanan} D., {Hamann} F., {Barlow} T., {Burbidge} E.~M., {Cohen} R.~D.,
  {Junkkarinen} V., {Lyons} R., 2004, \apj, 601, 715

\bibitem[{{Nestor} {et~al.}(2008){Nestor}, {Hamann}, \& {Hidalgo}}]{Nestor08}
{Nestor} D., {Hamann} F., {Hidalgo} P.~R., 2008, \mnras, 386, 2055

\bibitem[{{Peterson}(1997)}]{Peterson97}
{Peterson} B.~M., 1997, {An Introduction to Active Galactic Nuclei}. An
  introduction to active galactic nuclei, Publisher: Cambridge, New York
  Cambridge University Press, 1997 Physical description xvi, 238 p.~ISBN
  0521473489

\bibitem[{{Pounds} {et~al.}(2003){Pounds}, {King}, {Page}, \&
  {O'Brien}}]{Pounds03}
{Pounds} K.~A., {King} A.~R., {Page} K.~L., {O'Brien} P.~T., 2003, \mnras, 346,
  1025

\bibitem[{{Proga} \& {Kallman}(2004)}]{Proga04}
{Proga} D., {Kallman} T.~R., 2004, \apj, 616, 688

\bibitem[{{Proga} {et~al.}(2000){Proga}, {Stone}, \& {Kallman}}]{Proga00}
{Proga} D., {Stone} J.~M., {Kallman} T.~R., 2000, \apj, 543, 686

\bibitem[{{Reeves} {et~al.}(2009){Reeves}, {O'Brien}, {Braito}, {Behar},
  {Miller}, {Turner}, {Fabian}, {Kaspi}, {Mushotzky}, \& {Ward}}]{Reeves09}
{Reeves} J.~N., {O'Brien} P.~T., {Braito} V., {Behar} E., {Miller} L., {Turner}
  T.~J., {Fabian} A.~C., {Kaspi} S., {Mushotzky} R., {Ward} M., 2009, \apj,
  701, 493

\bibitem[{{Reichard} {et~al.}(2003){Reichard}, {Richards}, {Hall}, {Schneider},
  {Vanden Berk}, {Fan}, {York}, {Knapp}, \& {Brinkmann}}]{Reichard03}
{Reichard} T.~A., {Richards} G.~T., {Hall} P.~B., {Schneider} D.~P., {Vanden
  Berk} D.~E., {Fan} X., {York} D.~G., {Knapp} G.~R., {Brinkmann} J., 2003,
  \aj, 126, 2594

\bibitem[{{Rodriguez Hidalgo}(2009)}]{RodriguezHidalgo09}
{Rodriguez Hidalgo} P., 2009, PhD thesis, University of Florida, United States
  -- Florida

\bibitem[{{Sabra} \& {Hamann}(2005)}]{Sabra05}
{Sabra} B.~M., {Hamann} F., 2005, ArXiv Astrophysics e-prints

\bibitem[{{Sabra} {et~al.}(2003){Sabra}, {Hamann}, {Jannuzi}, {George}, \&
  {Shields}}]{Sabra03}
{Sabra} B.~M., {Hamann} F., {Jannuzi} B.~T., {George} I.~M., {Shields} J.~C.,
  2003, \apj, 590, 66

\bibitem[{{Shen} {et~al.}(2007){Shen}, {Strauss}, {Oguri}, {Hennawi}, {Fan},
  {Richards}, {Hall}, {Gunn}, {Schneider}, {Szalay}, {Thakar}, {Vanden Berk},
  {Anderson}, {Bahcall}, {Connolly}, \& {Knapp}}]{Shen07}
{Shen} Y., {Strauss} M.~A., {Oguri} M., {Hennawi} J.~F., {Fan} X., {Richards}
  G.~T., {Hall} P.~B., {Gunn} J.~E., {Schneider} D.~P., {Szalay} A.~S.,
  {Thakar} A.~R., {Vanden Berk} D.~E., {Anderson} S.~F., {Bahcall} N.~A.,
  {Connolly} A.~J., {Knapp} G.~R., 2007, \aj, 133, 2222

\bibitem[{{Silk} \& {Rees}(1998)}]{Silk98}
{Silk} J., {Rees} M.~J., 1998, \aap, 331, L1

\bibitem[{{Srianand} {et~al.}(2002){Srianand}, {Petitjean}, {Ledoux}, \&
  {Hazard}}]{Srianand02}
{Srianand} R., {Petitjean} P., {Ledoux} C., {Hazard} C., 2002, \mnras, 336, 753

\bibitem[{{Steffen} {et~al.}(2006){Steffen}, {Strateva}, {Brandt}, {Alexander},
  {Koekemoer}, {Lehmer}, {Schneider}, \& {Vignali}}]{Steffen06}
{Steffen} A.~T., {Strateva} I., {Brandt} W.~N., {Alexander} D.~M., {Koekemoer}
  A.~M., {Lehmer} B.~D., {Schneider} D.~P., {Vignali} C., 2006, \aj, 131, 2826

\bibitem[{{Telfer} {et~al.}(1998){Telfer}, {Kriss}, {Zheng}, {Davidsen}, \&
  {Green}}]{Telfer98}
{Telfer} R.~C., {Kriss} G.~A., {Zheng} W., {Davidsen} A.~F., {Green} R.~F.,
  1998, \apj, 509, 132

\bibitem[{{Trump} {et~al.}(2006){Trump}, {Hall}, {Reichard}, {Richards},
  {Schneider}, {Vanden Berk}, {Knapp}, {Anderson}, {Fan}, {Brinkman},
  {Kleinman}, \& {Nitta}}]{Trump06}
{Trump} J.~R., {Hall} P.~B., {Reichard} T.~A., {Richards} G.~T., {Schneider}
  D.~P., {Vanden Berk} D.~E., {Knapp} G.~R., {Anderson} S.~F., {Fan} X.,
  {Brinkman} J., {Kleinman} S.~J., {Nitta} A., 2006, \apjs, 165, 1

\bibitem[{{Turnshek}(1984)}]{Turnshek84}
{Turnshek} D.~A., 1984, \apj, 280, 51

\bibitem[{{Turnshek}(1988)}]{Turnshek88}
---, 1988, in Proceedings of the QSO Absorption Line Meeting, {Blades} J.~C.,
  {Turnshek} D.~A., {Norman} C.~A., eds., pp. 17--46

\bibitem[{{Turnshek} \& {Rao}(2002)}]{Turnshek02}
{Turnshek} D.~A., {Rao} S.~M., 2002, \apjl, 572, L7

\bibitem[{{Tytler} \& {Fan}(1992)}]{Tytler92}
{Tytler} D., {Fan} X.-M., 1992, \apjs, 79, 1

\bibitem[{{Vestergaard}(2003)}]{Vestergaard03}
{Vestergaard} M., 2003, \apj, 599, 116

\bibitem[{{Vestergaard} \& {Peterson}(2006)}]{Vestergaard06}
{Vestergaard} M., {Peterson} B.~M., 2006, \apj, 641, 689

\bibitem[{{Warner} {et~al.}(2003){Warner}, {Hamann}, \& {Dietrich}}]{Warner03}
{Warner} C., {Hamann} F., {Dietrich} M., 2003, \apj, 596, 72

\bibitem[{{Warner} {et~al.}(2004){Warner}, {Hamann}, \& {Dietrich}}]{Warner04}
---, 2004, \apj, 608, 136

\bibitem[{{Weymann} {et~al.}(1981){Weymann}, {Carswell}, \&
  {Smith}}]{Weymann81}
{Weymann} R.~J., {Carswell} R.~F., {Smith} M.~G., 1981, \araa, 19, 41

\bibitem[{{Weymann} {et~al.}(1991){Weymann}, {Morris}, {Foltz}, \&
  {Hewett}}]{Weymann91}
{Weymann} R.~J., {Morris} S.~L., {Foltz} C.~B., {Hewett} P.~C., 1991, \apj,
  373, 23

\bibitem[{{Yuan} {et~al.}(2002){Yuan}, {Green}, {Brotherton}, {Tripp},
  {Kaiser}, \& {Kriss}}]{Yuan02}
{Yuan} Q., {Green} R.~F., {Brotherton} M., {Tripp} T.~M., {Kaiser} M.~E.,
  {Kriss} G.~A., 2002, \apj, 575, 687

\end{thebibliography}

\label{lastpage}

\end{document}